\documentstyle[12pt]{article}
\begin{document}
\thispagestyle{empty}
\setlength{\baselineskip} {4.0ex}
\newcommand{\p}{$\overline{p}$}
\newcommand{\n}{$\overline{n}$}
\newcommand{\N}{$\overline{N}$}
\newcommand{\D}{$\overline{d}$}
\def\ie{{\it i.e.} }
\def\etal{{\it et al.}}
\def\p{$\pi$ }
\def\r{$\rho$ }
\def\o{$\omega$ }
\def\e{$\eta$ }
\def\s{$\sigma$ }
\def\ns{$N^*$ }
\def\en{$\eta$N }
\def\nn{$NN$ }
\def\nne{$NN \rightarrow NN \eta$ }
\def\ppe{$pp \rightarrow pp \eta$ }
\def\npe{$np \rightarrow np \eta$ }
\def\ppp{$pp \rightarrow pp \pi^0$ }
\def\de{$np \rightarrow d \eta$ }
\def\sa{$S/A$ }
\def\pt{$p_T$ }
\def\et{$E_T$ }
\def\fm3{fm$^3$}
\def\fmm3{fm$^{-3}$}
\def\fsi{$FSI\ $}
\def\isi{$ISI\ $}
\def\s11{$S11\ $}
\def\pr{Phys.\ Rev.\ }
\def\pra{Phys.\ Rev.\ }
\def\prb{Phys.\ Rev.\ }
\def\prc{Phys.\ Rev.\ }
\def\prd{Phys.\ Rev.\ }
\def\prl{Phys.\ Rev.\ Lett.\ }
\def\pl{Phys.\ Lett.\ }
\def\np{Nucl.\ Phys.\ }
\def\ap{Ann.\ Phys.\ (N.Y.) }
\def\prep{Phys.\ Rep.\ }
\def\jp{J.\ Phys.\ }
\def\zp{Z.\ Phys.\ }
\def\mpl{Mod.\ Phys.\ Lett.\ }
\def\rmp{Rev.\ Mod.\ Phys.\ }
\def\sjnp{Sov.\ J.\ Nucl.\ Phys.\ }
\title{{\hfill{\small{BGU PH-97/14}}}\protect\\ 
A Covariant OBE Model for $\eta$ Production in $NN$ Collisions}  
\author{ E.Gedalin\thanks{gedal@bgumail.bgu.ac.il},
A.Moalem\thanks{moalem @bgumail.bgu.ac.il} and L.Razdolskaja\thanks
{ljuba@bgumail.bgu.ac.il}}                                    
\date{Department of Physics \\
 Ben-Gurion University of the Negev\\
Beer-Sheva, 84105,  Israel
}

\maketitle
\clearpage
\begin{abstract}
A relativistic covariant one boson exchange model, previously 
applied to describe elastic nucleon-nucleon scattering, is extended 
to study $\eta$ production in NN collisions. The transition 
amplitudes for the elementary $BN \rightarrow \eta N$ processes with 
$B$ being the meson exchanged ( $B = \pi , \sigma , \eta , \rho , 
\omega$ and $\delta$) are taken to be the sum of four terms 
corresponding to  s and u-channels 
with a nucleon or a nucleon isobar N$^*$ (1535 MeV) in the intermediate
states. Taking the relative phases of the various exchange amplitudes 
to be +1, the model reproduces the cross sections for the 
$NN \rightarrow X \eta$ reactions in a consistent manner. In the limit 
where all $\eta$'s are produced via N$^*$ excitations, interference 
terms between the overall contribution from the exchange of 
pseudoscalar and scalar mesons with that of vector mesons cancel out. 
Consequently, much of the ambiguities in the model predictions due to 
unknown relative phases of different vector and pseudoscalar exchanges 
are strongly reduced.
\end{abstract}

\vfill
\pagebreak
\section{Introduction}
\bigskip

Much interest, both experimental[1-10] 
and theoretical[11-26], 
was devoted in recent years to the production of $\eta$ meson 
in hadronic collisions and electroproduction processes. In analogy 
with $\pi$ production through excitations of the $\Delta$ (1232 MeV) 
P33 resonance, it is believed that $\eta$-meson production in 
nucleon-nucleon (NN) collisions proceeds via the excitation of 
nucleon isobars. Because of isospin conservation however, only 
I=$\frac {1}{2}$ N$^*$ isobars which decay into $\eta$N pair are 
important. Pion and photon induced reactions\cite{krusche95a,brown79} 
provide strong evidence that near threshold, the 
N$^*$(1535 MeV) $S_{11}$ resonance dominates $\eta$ production. This is not
unexpected in view of the proximity of its mass 
to that of the $\eta$N pair and, its large branching
ratio ( 30-55$\%$ ) for N$^* \rightarrow \eta N$ decay\cite{pdg}.
The couplings to other resonances are week and seem to play a minor 
role close to threshold.

The production of $\eta$ mesons in NN collisions was considered 
by several groups[11-16] within one boson exchange (OBE) 
models, where the N$^*$ (1535 MeV) is excited via the exchange of a boson 
B and then 
decays into $\eta$N pair. Much of the uncertainties in the predictions 
of these model calculations concern the coupling of the mesons exchanged 
to the 
S11 resonance. The coupling of the $\pi$ and $\eta$ are known from 
the decay widths of this resonance but not so well the coupling of
other mesons. Assuming vector meson dominance,
Germond and Wilkin\cite{germond90} determined the $\rho$ coupling from 
$\gamma N \rightarrow N^*$ data and, as there is no evidence for 
isoscalar photon coupling to the $N^*$ they neglect the contribution
from $\omega$ exchange. Vetter et al.\cite{vetter91} determined the $\rho$
coupling from the empirical decay width of $5\%$ for
the N$^*$ to decay into two pions. They suggested a value 
$g_{\rho NN^*} = 0.615$ , nearly a factor of $\approx 2.7$ weaker 
than that deduced by Germond and Wilkin\cite{germond90}. Although 
no experimental evidence exists, they assumed that the $\omega$ also 
couples to the N$^*$ with $g_{\omega N N^*} = 0.436$, resulting with
about equal $\rho$ and $\omega$ contributions\cite{vetter91}.
Contrary to this, with the $\rho$ coupling 
taken from vector dominance, the cross section is dominated by 
$\rho$ exchange\cite{moalem96} while the $\omega$ 
with a coupling as proposed by Vetter et al.\cite{vetter91} may 
influence the total cross section at most by $\approx 30-40\%$. 
It is not clear though whether such a contribution
is required to explain the existing $pp \rightarrow pp \eta$ data. 

In addition to these uncertainties concerning the coupling constants,
the relative phases of different meson exchange amplitudes are 
unknown. Taking these to be ones of signs and restricting the model 
to $\pi$, $\eta$, $\omega$ and $\rho$ exchanges only, there are eight 
solutions for the transition amplitude, one for each of the sign 
combinations possible. A solution, where the $\rho$ and $\pi$ exchanges 
add destructively, seems to  explain best the cross section data for 
the $pp \rightarrow pp \eta$ reaction\cite{germond90,laget91,moalem96}. 

It is the aim of the present work to consider $\eta$ meson production
in NN collisions within a relativistic covariant OBE model,
where a virtual boson B produced on one of the incoming nucleons, is 
converted into an $\eta$ meson on the second nucleon (see Fig. 1). 
Each of the different exchange contributions to the production 
amplitude is determined by three factors, representing a source 
function of the B-N-N vertex, a boson propagator and a conversion 
amplitude for the $BN \rightarrow \eta N$ process. Only the latter 
depends on unknown boson-isobar couplings. 

In variation with previous calculations the $\eta$ production 
amplitude includes four terms, corresponding to s and u- channels 
with nucleon isobars (diagram 2a-2b) and nucleon excitations (diagram 
2c-2d) in the intermediate states. At threshold the contribution  
from s-channel N$^*$ pole (diagram 2c), hereinafter 
referred to as a resonance production term, dominates the process.
The other diagrams, by far less important, furnish a background term. 
Though the various exchange amplitudes and calculated 
cross sections remain similar to the ones calculated previously 
using a non relativistic OBE model\cite{moalem96}, the present model has 
distinct features which make predictions considerably more reliable. 
Particularly, in the limit where all $\eta$'s are produced via N$^*$ 
excitations, interference terms between the overall contribution 
from scalar and pseudoscalar meson exchanges with that from vector 
mesons cancel out. Consequently, the model predictions are insensitive 
to the choice of the relative phases of vector meson exchanges with 
respect to those of scalar and pseudoscalar mesons.

The paper is organized as follows. In Sect. 2 details of the 
formalism and model parameters are presented.   
In Sect. 3 we write the S-wave production amplitude in a simple
form suitable for numerical calculations. In Sect. 4 we
discuss interference between various exchange contributions. 
The relative importance of various contributions to the amplitude 
and cross section for the $NN \rightarrow NN \eta$ 
reactions is studied in Sect. 5. At energies 
close to threshold, final state interactions (FSI) strongly influence 
the scale and energy dependence of the calculated cross 
sections\cite{moalem95,moalem96}.  
These are introduced in Sect. 6 where  comparison with data is 
to be made. We conclude and summarize in Sect. 7.

\section{Formalism}
\bigskip
To calculate the primary production amplitude for the 
$NN \rightarrow X \eta$ reactions, with $X$ designating a bound or 
unbound two-nucleon state, we employ a fully covariant formalism based
on an effective OBE model. We assume a reaction mechanism as depicted
in Fig. 2, where a boson B created on one of the incoming nucleons is
converted into an $\eta$ meson on the second one via a conversion 
process, $BN \rightarrow \eta N$. We include both nucleon and N$^*$
(1535 MeV) isobar excitations in the intermediate states. 
We do not include diagrams in which the $\eta$ is produced on an internal 
meson line. Because of isospin and parity conservation only diagrams
with $\sigma$ and $\delta$ lines may contribute, but these
are  expected to be negligible small due to the weak coupling
constants involved.

\subsection{Lagrangian} The effective Lagrangian density is taken as, 
\begin{eqnarray}
  &  & L = \frac {f_{\pi NN}}{m_{\pi}} \bar{N}\gamma^5
       \gamma^{\mu}\partial_{\mu} \vec{\pi} \vec{\tau}N +   
       \frac {f_{\eta NN}}{m_{\eta}}\bar{N}\gamma^5 
       \gamma^{\mu}\partial_{\mu}\eta  N + 
        g_{\sigma NN}\bar{N} \sigma N + \nonumber \\ 
  &  &     g_{\delta NN}\bar{N}\vec{\tau}\vec{\delta} N + 
 	\bar{N}[g_{\rho NN}\gamma^\mu +
 	\frac{f_{\rho NN}}{2M}\sigma^{\mu\nu}\partial_{\nu}]
       \vec{\tau} \vec{\rho}_\mu N +
        g_{\omega NN} \bar{N}\gamma^\mu \omega_\mu N + \nonumber \\
  &  &     [ g_{\pi NN^*}\bar{N}^*\vec{\tau} \vec{\pi} N +
        g_{\eta NN^*}\bar{N}^* \eta N + 
        i g_{\sigma NN^*}\bar{N}^*\gamma^5 \sigma N + \nonumber \\
  &  &     g_{\rho NN^*}\bar{N}^*\gamma^5\gamma^\mu 
        \vec{\tau} \vec{\rho}_\mu N + 
        g_{\omega NN^*} \bar{N}^*\gamma^5\gamma^\mu \omega_\mu N + 
        i g_{\delta NN^*}\bar{N}^*\gamma^5 \delta N +  h.c.]~, ~~
\label{lagran} 
 	\end{eqnarray}
where N, N$^*$, $\pi ,\sigma ,\eta ,\rho ,\omega$ and $\delta$
represent  the fields of a nucleon, a nucleon isobar and the
mesons  in the spin isospin space; 
M and $m_B$ are the the mass of the nucleon 
and a meson B; $\vec{\tau}$ are Pauli matrices 
acting in the isospin space and the 
$\gamma$'s denote Dirac matrices. 
This is a simple generalization of a Lagrangian  
used by Machleidt et al.\cite{machleidt89} to describe NN elastic 
scattering data. As in Ref.\cite{machleidt89} a pseudovector 
coupling is assumed for the $\pi$NN and $\eta$NN vertices and the
$\omega$NN tensor coupling $f_{\omega NN}$ is set to be zero.
It is to be noted that using pseudoscalar couplings rather than 
pseudovector couplings for the $\pi$ and $\eta$ mesons would not alter the 
principal conclusions from the present work. But, taking the $\omega$
tensor coupling to be zero has the consequence that the 
$\omega$ exchange contribution to $\eta$ production rate via N$^*$ 
excitations  vanishes for the $pp \rightarrow pp \eta$ reaction
(see Sect.  3). 

Note also that the interaction Lagrangian densities 
for the $\sigma$ and $\delta$ 
mesons in Eqn. \ref{lagran} are chosen in the form
$i g_{B NN^*} [\bar{N}^*\gamma^5N B + \bar{N}\gamma^5N^* B]$.
As we show in Appendix A, the coupling constants $g_{B NN^*}$ 
($ B=\sigma , \delta $) are related through an effective triangle 
diagram, to $g_{B NN} , g_{\pi NN}$ and $g_{\pi NN^*}$,
and this form ascertains that all of these 
quantities be real numbers.

\subsection{The primary production amplitudes}
Suppose now that the production of a pseudoscalar meson 
$NN \to NN P$ reaction proceeds via the mechanism depicted in Fig. 1.
The primary production amplitude can then be written in the form,
\begin{equation}
M^{in}_{if} = 
\sum_{B}[T_{BN\rightarrow PN}(p_4,k;p_2,q)G_B(q)S_{BNN}(p_3,p_1)]
 +[1\leftrightarrow	2;3\leftrightarrow 4],~~
	\label{matrel}
\end{equation}
where $S_{BNN}(p_3,p_1)$ and $G_B(q)$ represent the  source function
and propagator of the boson B,
and $T_{BN\rightarrow PN}$ being the amplitude for the 
$BN\rightarrow PN$ transition. For the reactions to be considered,
the latter depends on  the unknown couplings to the nucleon isobar. 
The parameterization of the source functions and boson propagators
is rather well determined from fitting NN elastic scattering data.
In Eqn. \ref{matrel} 
$p_i,q$ and $k$ designate 4-momenta for the i-th 
nucleon, the boson exchanged (B) and the meson produced (P). 
The sum runs over all possible boson exchanges which contribute to 
the process. The bracket $[1\leftrightarrow 2;3\leftrightarrow 4]$ 
stands for a similar sum with the momenta $p_1, p_3$ and $p_2,p_4$ 
being interchanged. 

To make the present paper reasonably self contained we write in 
what follows the expressions for the source functions, 
propagators and conversion amplitudes. We use covariant 
parameterizations for $S_{BNN}$ and $T_{BN \rightarrow PN}$. 
For scalar and pseudoscalar mesons the propagator and source 
function are\cite{gaziorovitch69,IZ},
\begin{eqnarray}
 & & G_{S,P}(q^2) = i/(q^2 - m^2_B +i\epsilon)~,
 \label{propsp} \\
 & & S_{SNN} = \bar{u}(p_3) Iu(p_1)F_S(q^2)~,
\label{ssource} \\
 & & S_{PNN} = \bar{u}(p_3)\gamma^5 Iu(p_1)F_P(q^2)~,
\label{psource}
\end{eqnarray}
where $F_B(q^2)$ is a source form factor and $I$ the appropriate isospin 
operator, i.e., $I=1$ and $I=\vec{\tau}$ for isoscalar and 
isovector mesons, respectively. 
Here $u$ is a nucleon Dirac spinor and $p_3 = p_1 - q$ is 
the final nucleon momentum (see Fig. 1). 
The Dirac spinors as well as the amplitudes
$T_{BN \rightarrow NN}$ and $M_{if}^{in}$ are normalized as in
Itzykson and Zuber\cite{IZ}.

Without making unnecessary limiting assumptions about the reaction 
mechanism, the amplitude for the conversion process of a scalar 
into a pseudoscalar meson , $SN\rightarrow PN$, has the usual 
form \cite{gaziorovitch69},
\begin{equation}
T_{SN\rightarrow PN}(p_4,k;p_2,q) = \bar{u}(p_4)\gamma^5[A_S
+\frac {k\! \! \! / + q\! \! \! / }{2}B_S ]Iu(p_2)~.
\label{tsp}
\end{equation}
Similarly, the amplitude for the conversion of a pseudoscalar 
meson into another  pseudoscalar meson, $P_1N\rightarrow P_2N$, 
is given by,
\begin{equation}
T_{P_1N\rightarrow P_2N}(p_4,k;p_2,q) = \bar{u}(p_4)[A_P
+\frac{1}{2}(k\! \! \! /  + q\! \! \! / )B_P ]Iu(p_2)~.
\label{tpp}
\end{equation}
Here  $a\! \! \! /  = \gamma^{\mu} a_{\mu}$ and the
quantities $A_{S(P)}$, $B_{S(P)}$ are invariant functions depending
on the Mandelstam variables. These functions have 
an isospin structure in accordance with the following rules :\\
i)$A_S = A \times  1$ for both B and P being isoscalar particles;\\
ii)$A_S = A\times\vec{\tau}$ for B being isovector and P isoscalar 
or vice versa;\\
iii)$A^{ab}_S = 
A^0\times\delta^{ab} + A^1\times[\tau^a, \tau^b]$ for B and P being 
both isovector particles.
In Subsection. 2.3 we shall calculate these functions for the model 
proposed using the effective interaction Lagrangian, Eqn. \ref{lagran}.
 
The propagator for a vector meson is defined as,
\begin{equation}
 G^V_{\mu\nu} = -i \frac {g_{\mu\nu} - q_\mu q_\nu/ m^2} 
{q^2  - m^2_B +i\epsilon}~,
\label{propv}
\end{equation}
where $g_{\mu\nu}$ is the metric tensor.
Generally, a vector source is the sum of vector and tensor current terms,
 \begin{equation}
S^{\mu}_{VNN}(p_1,p_3) = \bar{u}(p_3)[\gamma^\mu F^{(1)}_V(q^2) 
+ i\sigma^{\mu\nu} q_{\nu} F^{(2)}_V(q^2)] Iu(p_2)~,
\label{vsource}
 \end{equation}
with $F^{(1)}_V(q^2)$ and $F^{(2)}_V(q^2)$ standing for vector and
tensor source form factors, quantities being the analogous of the nucleon 
electromagnetic form factors. This  last expression contains conserved 
currents only and  therefore satisfies current conservation, $i.e.$,
\begin{equation}
q_\mu S^\mu_VNN(q) = 0~.
\label{ccon}
\end{equation} 
Likewise, the conversion amplitude of a vector meson into 
a pseudoscalar meson, $VN \rightarrow P N$, is expressed in terms 
of eight invariant functions $A_{V,i} ( i=1 - 8)$ all having
an isospin structure as  mentioned above for the $A_S$ and $B_S$.
More concisely we write, 
\begin{equation}
T^{\mu}_{V N\rightarrow PN}(p_4,k;p_2,q) = \bar{u}(p_4)\gamma^5
[\gamma^\mu A^V_{12}
+ p_4^\mu A^V_{34} + k^\mu A^V_{56} + q^\mu A^V_{78}] Iu(p_2)~,
\label{tvp}
\end{equation}
where $A^V_{ij} = A_{V,i} - k\! \! \! / A_{V,j}$. It is worth noting 
that because of current conservation, Eqn. \ref{ccon},
only the first term of Eqn. 8 may contribute to the production process
so that,
\begin{equation}
G_{\mu \nu}^V S_{VNN}^{\nu} \equiv \frac {-i g_{\mu\nu}}{q^2 - m_B^2 + 
i \epsilon } S_{VNN}^{\nu}~.
\end{equation}
Also, for this same reason, the term $q^{\mu} A_{78}^V$ in Eqn. \ref{tvp} 
does not contribute to the production amplitude, $M^{(in)}$.
By using straightforward algebra of $\gamma$ matrices and the Dirac 
equation for a free particle and, by substituting equivalent 
two-component free spinor matrix elements (Table A6.1 of Ref. 
\cite{ericson}) in Eqns. \ref{propsp}-\ref{tvp} we may write the 
primary production amplitude as, 
\begin{equation}
M^{in}_{NN\rightarrow PNN}(p_1,p_2;p_3,p_4,k) = \sum_{S}
M^{in}_S + \sum_{P} M^{in}_P + \sum_{V} M^{in}_V
+[1\leftrightarrow	2;3\leftrightarrow 4],
\label{tranamp}
\end{equation}
where the sums over S, P, and V run, respectively, over all 
scalar, pseudoscalar and vector mesons and,
\begin{eqnarray}
 & & M^{in}_S = S_0 + S_{13}^k \sigma_{13}^k + \tilde{S}_{24}^k 
\sigma_{24}^k
             +  S^{kl} \sigma_{13}^k \sigma_{24}^l~, \label{sexampl}\\
 & & M^{in}_P = P_{13}^k \sigma_{13}^k  
             +  P^{kl} \sigma_{13}^k \sigma_{24}^l~, \label{pexampl} \\
 & & M^{in}_V = V_0 + V_{13}^k \sigma_{13}^k + \tilde{V}_{24}^k 
\sigma_{24}^k
             +  V^{kl} \sigma_{13}^k \sigma_{24}^l~~. \label{vexampl}
\end{eqnarray}
In Eqns. \ref{sexampl}-\ref{vexampl} all of the quantities $S_0, V_0$; 
$S_{ij}, P_{ij}, V_{ij}$; $\tilde{S}_{ij}$, 
$\tilde{V}_{ij}$; $S^{nm}, P^{nm}, V^{nm}$ are functions of
the particle momenta involved. Their definitions are given 
in Appendix B for the general case in terms of the invariant functions.

\subsection{The invariant functions}
To complete the derivation of the $\eta$ production amplitude the 
invariant functions $A$ and $B$ must be specified. To do this 
we assume that the conversion process proceeds via the mechanisms of Fig. 3,
and consider as an example the transition amplitude for the 
$\omega N \rightarrow \eta N$ process. Using the usual Feynman 
rules and vertices of the Lagrangian, Eqn. \ref{lagran}, one writes,
\begin{eqnarray}
 & & T^\mu_{\omega p\rightarrow \eta p}=
 - i u(p_4) \{ g_{\omega NN^*} g_{\eta NN^*} \gamma^\mu \gamma^5
[ G_R (s_{13}) + G_R (u_{13}) ] +
\nonumber \\
 & & g_{\omega NN} \frac {f_{\eta NN}}{m_{\eta}} 
[k\! \! \! / \gamma^\mu \gamma^5 G_N (s_{13}) - 
\gamma^5 (k_\mu-i\sigma^{\mu\nu}k_\nu) G_N (u_{13}) ]\} u(p_2)~,
\label{omgcampl}    
\end{eqnarray}
where $s_{13} = (k + p_4)^2$, $u_{13} = (p_2 - k)^2$ and $ G_N $ 
denotes a nucleon propagator
\begin{equation}
 G_N =  \frac { i }{ p\! \! \! / - M + i \epsilon}~.
\label{propf}
\end{equation}
For an isobar the mass in Eqn. \ref{propf} is
replaced by a mass operator, $i.e.$,
 \begin{equation}
G_R = \frac{i} {p\! \! \! /  - M_R + i \Gamma_R / 2 }~.
\label{rprop}
\end{equation}
Generally speaking, $M_R$ and $\Gamma_R$ are the real and imaginary
parts of the isobar self energy, quantities representing the mass 
and width of the resonance. Both of these are functions of 
the Mandelstam variable $s=p^2$. It is straightforward to show 
that at resonance, where $ \sqrt {s} \approx M_R $, 
Eqn. \ref{rprop} simplifies to the usual Breit-Wigner form, 
\begin{equation}
G_R (s) = \frac {-i} {M_R - \sqrt{s} - i \Gamma_R / 2 }~.
\label{rsprop}
\end{equation}
For future use we define two related quantities,
\begin{equation}
\Delta_N(x) = i/(x-M^2+i\epsilon)~,
\end{equation}
and
\begin{equation}
\Delta_R(x) = -i/(M_R-\sqrt{x}-i\Gamma_R/2)~.
\end{equation}
Altogether there are four terms in Eqn. \ref{omgcampl} corresponding 
to s and u-channels with N$^*$ (diagrams 3a-3b) and with nucleon 
(diagrams 3c-3d) poles. 
By comparing Eqn. \ref{omgcampl} with Eqn. \ref{tvp} one obtains,
\begin{eqnarray}
 & & A_{\omega,1} =
g_{\omega NN^*}g_{\eta NN^*} \tilde{G}_R
+ ig_{\omega NN}\frac{f_{\eta NN}}{m_{\eta}} X_V ~,
\label{ao1} \\
 & & A_{\omega,2} =
 -2iMg_{\omega NN}\frac{f_{\eta NN}}{m_{\eta}}\tilde{G}_N ~, 
\label{ao2} \\
 & & A_{\omega,5} =
 i2Mg_{\omega NN}\frac{f_{\eta NN}}{m_{\eta}}\Delta_N(s_{13}) ~,
 \label{ao5} \\
 & & A_{\omega,3} = A_{\omega,4} = A_{\omega,6} = A_{\omega,7}
              = A_{\omega,8}=0~.
\end{eqnarray}	
Similar calculations for other meson exchanges give the following:
\begin{eqnarray}
 & & A_\sigma =  g_{\sigma NN^*}g_{\eta NN^*} \tilde{G}_R
    - ig_{\sigma NN}\frac{f_{\eta NN}}{m_{\eta}} X_S~,\label{as} \\
 & & B_\sigma = i 2Mg_{\sigma NN}\frac{f_{\eta NN}}
{m_{\eta}}\tilde{G}_N~, \label{bs} \\
 & & A_\delta = g_{\delta NN^*}g_{\eta NN^*} \tilde{G}_R\vec{\tau}
 - i g_{\delta NN}\frac{f_{\eta NN}}{m_{\eta}} X_S\vec{\tau} ~,\label{ad} \\
 & & B_\delta = i 2Mg_{\delta NN}\frac{f_{\eta NN}}{m_{\eta}}
 \tilde{G}_N \vec{\tau}  ~, \label{bd} \\
 & & A_\pi = g_{\pi NN^*}g_{\eta NN^*} \tilde{G}_R\vec{\tau} ~,\label{api} \\
 & & B_\pi = -\frac{f_{\eta NN}}{m_{\eta}}\frac{f_{\pi NN}}{m_{\pi}}
  X_P\vec{\tau} ~,\label{bpi} \\
 & & A_\eta = g^2_{\eta NN^*} \tilde{G}_R ~,\label{aeta} \\
 & & B_\eta = -\frac{f^2_{\eta NN}}{m^2_{\eta}}X_P ~, \label{beta} \\
 & & A_{\rho,1} = g_{\rho NN^*}g_{\eta NN^*} \tilde{G}_R\vec{\tau}
 + ig_{\rho NN}\frac{f_{\eta NN}}{m_{\eta}}X_V\vec{\tau}~,\label{ar1}\\
 & & A_{\rho,2} =-2iMg_{\rho NN}\frac{f_{\eta NN}}{m_{\eta}}Y_V\vec{\tau}~,  
\label{ar2} \\
 & & A_{\rho,3} = - ig_{\rho NN}\frac{f_{\eta NN}}{m_{\eta}}
 \frac{\kappa}{2M}Z_V \vec{\tau}~, \label{ar3} \\
 & & A_{\rho,4} = - 2ig_{\rho NN}\frac{f_{\eta NN}}{m_{\eta}}
 \kappa \tilde{G}_N  \vec{\tau}~,\label{ar4} \\
 & & A_{\rho,5} = i2Mg_{\rho NN}\frac{f_{\eta NN}}{m_{\eta}}
 (1 + 2\kappa)\Delta_N(s_{13})\vec{\tau}~, \label{ar5} \\
 & & A_{\rho,6} = -2ig_{\rho NN}
 \frac{f_{\eta NN}}{m_{\eta}}\kappa \Delta_N(s_{13}) \vec{\tau}~.\label{ar6}
\end{eqnarray}
In Eqns. \ref{ao1} - \ref{ar6} we  use the notations,
\begin{eqnarray}
 & & \tilde{G}_R = \Delta_R(s_{13}) +  \Delta_R(u_{13})~,\\
 & & \tilde{G}_N = \Delta_N(s_{13}) +  \Delta_N(u_{13})~, \\
 & & X_S = (2p_2k + m^2_\eta - 2M^2)\Delta_N(s_{13}) + 
     (m^2_\eta -2p_1k - 2M^2)\Delta_N(u_{13})~,~~~~ \\
 & & X_P = (2p_2k + m^2_\eta + 4M^2)\Delta_N(s_{13}) + 
     (m^2_\eta -2p_1k + 4M^2)\Delta_N(u_{13})~,~~~~ \\
 & & X_V = (2p_2k+m^2_\eta)\Delta_N(s_{13}) + 
	  (2p_1k+m^2_\eta) \Delta_N(u_{13})~,\\
 & & Y_V = (1+\kappa + 
     \frac{\kappa}{4M^2}(2p_2k+m^2_{\eta}))\Delta_N(s_{13}) + \nonumber\\
 & & ~~~~~~(1+\kappa + \frac{\kappa}{4M^2}(2p_1k-m^2_{\eta})) 
           \Delta_N(u_{13})~,\\
 & & Z_V = (2p_2\cdot k + m^2_{\eta})\Delta_N(s_{13}) 
	 - (2p_1\cdot k - m^2_{\eta})\Delta_N(u_{13})~. 
\end{eqnarray}
\subsection{The model parameters}
In the calculations to be presented below all of the 
meson-nucleon-nucleon couplings are taken from  Machleidt 
et al.\cite{machleidt89}; their relativistic OBEP set with pseudovector 
coupling for the pseudoscalar mesons. For the sake of completeness these 
are included in Table \ref{mesonpar}.
As in Ref.\cite{machleidt89} the source form factors are taken to be,
$$F_{\sigma} = g_{\sigma NN}f_{\sigma}~,~~~
F_{\pi} = ig_{\pi NN}f_{\pi}~,~~~ 
F_{\eta} = ig_{\eta NN}f_{\eta }~,~~~ 
F_{\delta} = g_{\delta NN} f_{\delta}$$
$$F_{\rho}^{(1)} = g_{\rho NN}f_{\rho}~,~~~  
F_{\rho}^{(2)} = \frac{\kappa}{2M}g_{\rho NN}f_{\rho }~,~~~
F_{\omega}^{(1)} = g_{\omega NN}f_{\omega }~,$$
with the parameterization,
\begin{equation}
f_{B }(q^2)=\frac{\Lambda^2_{B}-m^2_B}
{\Lambda^2_{B}-q^2}~.
\label{formf}
\end{equation}	
The $\pi$ and $\eta$ couplings to the $N^*$ (1535 MeV)
are deduced from
the partial widths of the N$^*$ to decay into $\pi$N 
and $\eta$N; their values are rather well accepted (see 
Refs. [12-16]). The $g_{\rho NN^*}$ is deduced 
from $\gamma N \rightarrow N^*$ data and vector dominance\cite{germond90}. 
For the other mesons there exits no direct information which 
can be used to determine their coupling to the N$^*$. In 
Ref.\cite{vetter91} the $\omega$ coupling is evaluated from the relation,
\begin{equation}
\frac{g_{\omega NN^*}}{ g_{\omega NN}} = \frac{g_{\pi NN^*}}{ g_{\pi NN}}~.
\label{vet}
\end{equation}
Assuming SU(3) flavor symmetry it can be shown also that,
\begin{equation}
\frac{g_{\omega NN^*}}{ g_{\omega NN}} = \frac{g_{\rho NN^*}}{ g_{\rho NN}
(1 + \kappa_{\rho})}~.	
\label{su3flv}
\end{equation}
Taking the appropriate constants from Table \ref{mesonpar} one 
obtains $g_{\omega NN^*} \approx 1$ from both of these expressions.
This is a factor of two higher than that suggested by Vetter et 
al.\cite{vetter91}. 
Similarly, from using effective triangle diagrams 
(see Appendix A)  the $g_{\sigma NN*}$ and $g_{\delta NN^*}$ are 
also related to the pion constants through an expression similar 
to Eqn. \ref{vet}.

\section{Amplitudes and cross sections}
In this section we write S-wave production amplitudes
for the $pp \rightarrow pp \eta$ and $pn \rightarrow pn \eta$ reactions 
in a simple form which is suitable for numerical calculations. To this 
aim let us call,
\begin{equation}
\Pi_j = \frac {{\bf p}_j} {\sqrt{E_j + M}}~,\\
\end{equation}
where ${\bf p}_j$ and $E_j$ are three-momentum and total energy 
of the j-th nucleon. For the incoming particles in the CM system,
$\Pi_1 = - \Pi_2 = \Pi$. The total energy square is $s = (p_1 + p_2)^2$.
Then the energy available in the CM system is,
\begin{equation}
Q = \sqrt{s}- 2M - m_\eta~~.
\end{equation} 
We shall calculate the amplitudes and cross sections as  functions 
of $Q$. 

There are two amplitudes one isovector $M_{11}$ and one isoscalar 
$M_{00}$ which determine completely the cross sections for the 
$NN \rightarrow X \eta$ reactions at rest. These two amplitudes 
correspond to the ${}^{33}P_0 \rightarrow {}^{31}S_0$ and 
${}^{11}P_1 \rightarrow {}^{31}S_1$ transitions in the two-nucleon 
system. Only  $M_{11}$ contributes to the rate of the 
$pp \rightarrow pp \eta$ reaction. Likewise only $M_{00}$ contributes 
to the $np \rightarrow d \eta$ reaction, but both amplitudes contribute 
to the rate of the $np \rightarrow np \eta$ reaction. 
Following the discussion in subsection 2.2 we write these amplitudes
as,
\begin{eqnarray}
& & M_{11} = M_{\pi} + M_{\eta} + M_{\sigma} + M_{\delta} + M_{\rho} 
       + M_{\omega}~,\\
& & M_{00} = -3 M_{\pi} + M_{\eta} + M_{\sigma} - 3 M_{\delta} 
       -3 M_{\rho} + M_{\omega}~,
\label{matone}
\end{eqnarray}
where $M_{B}$ stands for a partial exchange amplitude of 
a meson B. The factor of $-3$ is due to isospin.
By evaluating matrix elements of the expressions, Eqns. 14-15, 
between the allowed initial and final spin states and 
substituting for the appropriate 
invariant functions listed in subsection 2.3, we obtain,
\begin{eqnarray}
 & & M_{\sigma} = i G_{\sigma NN} (g_{\sigma NN*}g_{\eta NN*} R + 
g_{\sigma NN}g_{\eta NN}\Sigma_S)~, \label{minsigma} \\
 & &  M_{\delta} = i G_{\delta NN} (g_{\delta NN*}g_{\eta NN*} R + 
g_{\delta NN}g_{\eta NN}\Sigma_S)~, \label{mindelta} \\
 & & M_{\pi}= i G_{\pi NN} (g_{\pi NN*}g_{\eta NN*} R + 
g_{\pi NN}g_{\eta NN}\Sigma_P)~, \label{minpi} \\
 & & M_{\eta} = i G_{\eta NN} (g^2_{\eta NN*} R + 
g^2_{\eta NN}\Sigma_P]~, \label{mineta} \\
 & & M_\rho = G_{\rho NN} [g_{\rho NN*}g_{\eta NN*}
(w_{\rho} \pm 2 v_{\rho}) R + 
g_{\rho NN}g_{\eta NN}(\Sigma_{\rho}^{(1)} + 2 \Sigma_{\rho}^{(2)})]~,~~
\label{minrhou} \\
 & & M_\omega = G_{\omega NN} [g_{\omega NN*}g_{\eta NN*}
(w_{\omega} \pm 2 v_{\omega}) R + 
g_{\omega NN}g_{\eta NN}(\Sigma_{\omega}^{(1)} + 2 \Sigma_{\omega}^{(2)})]~.
~~~~~\label{minomega} 
\end{eqnarray}
In Eqns. 55-60 we have used the notation,
\begin{eqnarray}
 & & G_{BNN}= g_{B NN}\frac{E + M}{M}\frac{1}{M(m_{\eta}+Q)
+ m^2_{\pi}}f^2_{\sigma}(-M[m_{\eta}+Q]) \Pi~,\\	 
 & & R = \frac{1}{M_R-M-m_{\eta}-Q-i\Gamma/2}
+ \frac{1}{M_R-M+m_{\eta}+Q-i\Gamma/2}~,~~~~~~\\
 & & \Sigma_S =\frac{1}{M}\left(1 - \frac{5m_\eta}
{2M-m_\eta}\right)~,\\
 & & \Sigma_P = \left(\frac{m_{\eta}}{2M}\right)^2
\left[1 + \frac{2(E +M)}{m_{\eta}}{ 
\Pi \cdot \Pi}\right]\frac{1}{2M+m_\eta}~.\\ 
 & & \Sigma_{\rho}^{(1)} = i \frac{m_\eta}{4M^2}\nonumber\\
 & & ~~~~~~\left\{(1+\kappa \Pi \cdot \Pi)\left[2-(1-
\frac{\kappa}{2})\frac{m_\eta}{2M+m_\eta}\right] +
\frac{\kappa m^2_\eta}{4M(2M+m_\eta)}\right\}~, \\
 & & \Sigma_{\rho}^{(2)} = i\kappa (1+\kappa)\frac{m_\eta}{2M}
\nonumber \\
 & &  ~~~~~~\left\{-\frac{E+M}{M} {\Pi \cdot \Pi}
\left(1-\frac{m_\eta}{2M+m_\eta}\right) +
\frac{m^2_\eta}{4M(2M+m_\eta)}\right\}~,\\
 & & \Sigma_{\omega}^{(1)}  
= i\frac{m_\eta}{4M^2}\left(2-\frac{m_\eta}{2M+m_\eta}\right)~,\\
 & & \Sigma_{\omega}^{(2)} = 0~, \\
 & & w_V = 2 + \kappa_V \left(\frac{E+M}{2M}\right) {\Pi \cdot \Pi}~,  \\
 & & v_V = 1 + \kappa_V \left(\frac{E+M}{2M}\right)~.\\
\end{eqnarray}
The $-$ and $+$ signs in the expressions for $M_{\rho}$ and $M_{\omega}$ 
refer to the ${}^{33}P_0 \rightarrow {}^{31}S_0$ and 
${}^{11}P_1 \rightarrow {}^{13}S_1$ transitions, respectively. 
The expressions for other
exchange amplitudes are the same for both of these transitions. 
Because of kinematic, the resonance terms in Eqns. 55-60 exceed by
far any of the background terms.
Following the definitions of $w_{\rho}$ and $ v_{\rho}$, 
the $\rho$ exchange amplitude, 
Eqn. 59, has opposite signs for the two transitions but,
because of the negative isospin factor the $\rho$ contributions in
$M_{11}$ and $M_{00}$ have equal signs.
Also note that for $\kappa_{\omega} \equiv 0$, $w_{\omega}$=2 and 
$v_{\omega}$=1 causing the first term in Eqn. \ref{minomega} to vanish
in the case of a ${}^{33}P_0 \rightarrow {}^{31}S_0$ transition.
Thus, only small background terms may contribute to the $\omega$ 
exchange amplitude, and practically, the calculated cross section for the 
$pp \rightarrow pp \eta$ reaction is expected to be free from
ambiguities  due to the $\omega$NN$^*$ coupling.

Finally a general formula for the invariant cross section  for the 
$NN \to NN \eta$ reaction is written as
\begin{equation}
\sigma_{TT} = \frac {{(2\ \pi )}^{-5}}{128 \ p^* s^{3/2}}
\int  d\Omega_1 \ d\Omega_2 \ \frac {d\sqrt {s_{23}}} {\sqrt {s_{23}}} 
\lambda^{1/2} (s, m^2_1, s_{23}) \ 
\lambda^{1/2} (s_{23}, m^2_2, m^2_3) \ 
\ |\ M_{TT}\  Z_{}\ |^2 \ \ ,
\label{eq:14}
\end{equation}
where $M_{TT}$ represents the amplitudes $M_{11}$ and $M_{00}$ of Eqns. 
53-54; the indices 1, 2 and 3 label the outgoing particles, 
$\sqrt {s}$ and $\sqrt {s_{23}}$ are  the total energy 
and partial energy of particles 2 and 3, 
$ p^*$ is the center 
of mass (CM) momentum of the incoming proton, and
$ \ Z_{}\ $ is the FSI correction factor to be specified below.   
The $\lambda $ is the usual triangle function defined as\cite{byckling}
\begin{equation}
\lambda (x, y, z) = x^2\ +\ y^2\ +\ z^2\ -\ 2xy\ -\ 2xz\ -\ 2yz\ \ .
\label{eq:15}
\end{equation}

\section{Interference terms}
As indicated in the introduction, major uncertainties in 
previous model calculations arise from the unknown relative 
phases of different meson contributions. Particularly, 
interference terms between prominent $\rho$ and $\pi$
contributions
influence strongly the scale of the cross section. The aim of the 
present section is to show that in the present model and 
in the limit where $\eta$-meson 
production proceeds through N$^*$ (1535 MeV) excitations only, 
interference terms between the overall contribution from scalar 
and pseudoscalar meson exchanges and that of
vector meson exchanges cancel out. To see this let us 
consider the contributions to the production amplitude from 
N$^*$ pole terms only. By substituting the various exchange 
contributions, Eqns. 55-60, into Eqn. 2 one obtains,
\begin{eqnarray}
   & & {M}^{in}_{NN\rightarrow PNN} (p_1, p_2; p_3, p_4, k) = 
g_{\eta NN^*} R  \nonumber \\
   & & {\large [} i ( G_{\pi NN} g_{\pi NN^*} + 
G_{\eta NN} g_{\eta NN^*} +G_{\sigma NN} g_{\sigma NN^*} + 
G_{\delta NN} g_{\delta NN^*} ) + \nonumber \\
   & &  G_{\rho NN} g_{\rho NN^*} ( w_{\rho} \pm 2 v_{\rho}) +
G_{\rho NN} g_{\rho NN^*} ( w_{\omega} \pm 2 v_{\omega})  {\large ]} ~.
\label{nstex}
\end{eqnarray}
We notice that all of the quantities $G_{BNN}, g_{BNN^*}, w_V $ and $v_V$
are real so that 
\begin{eqnarray}
  & & M^{(in)} M^{(in)\dag} = 
| g_{\eta NN^*} R |^2  \nonumber \\
   & & {\large \{}  [ G_{\pi NN} g_{\pi NN^*} + 
                     G_{\eta NN} g_{\eta NN^*} + 
                     G_{\sigma NN} g_{\sigma NN^*} + 
        G_{\delta NN} g_{\delta NN^*} ]^2 + \nonumber \\
   & & \left[ G_{\rho NN} g_{\rho NN^*} ( w_{\rho} \pm 2 v_{\rho}) +
       G_{\omega NN} g_{\omega NN^*} ( w_{\omega} \pm 2 v_{\omega}) 
\right]^2  {\large \}} ~.
\label{nstex}
\end{eqnarray}
We may thus conclude that in a model where isobar mechanism dominates,
in practice, the $\pi$ and $\rho$ exchange amplitudes add incoherently
and their unknown relative phases should introduce no ambiguities in
the model predictions.  There remain of course ambiguities due 
to interference amongst the different vector meson exchanges or
amongst the various scalar and pseudoscalar meson exchanges to be
studied below in more details.

\section{Predictions}
We now apply the model presented in the previous sections to calculate the 
energy integrated cross section for the $pp \rightarrow pp \eta$ 
and $pn \rightarrow pn \eta$ reactions. 
We consider first the relative importance of 
various exchange contributions. To this
aim we draw in Figs. 4-5 the partial exchange amplitudes, Eqns. 55-60.
Most important are the $\rho$ and $\pi$ exchanges.
Other contributions are small but might be influential through 
interference. 
Taking the relative phases of the different exchange amplitudes 
to be ones of signs, there are altogether 32 solutions, one 
for each of the different sign combinations. All of these 
solutions yield cross sections having identical energy 
dependence but vary in scale. 
The primary production amplitudes for the $pp \rightarrow pp \eta$
and $pn \rightarrow pn \eta$ reactions calculated with all
of the relative phases chosen to be +1 are shown in Fig.6. 
As we shall see below this phase combination, hereinafter 
referred to as the standard solution, explains best the cross section 
data at energies close to threshold for the $pp \rightarrow pp \eta$, 
$pn \rightarrow pn \eta$. The cross sections corresponding to
these amplitudes are drawn as solid lines in Figs. 7-8. The other 
curves in these figures 
represent predictions with other phase combinations, representing 
the lowest (dashed curve) and highest (dotted curve) cross sections 
obtained. At most the scale of the calculated cross sections 
varies by a factor of $2-3$, as compared to a factor of about 20 
reported for other models\cite{moalem96}.
To examine these ambiguities in the model predictions let us 
consider the standard solution in some details. Consider first the 
$pp \rightarrow pp \eta$ reaction. Here the individual contributions 
from all of the scalar and pseudoscalar meson exchanges have equal 
signs and therefore should add constructively. 
In Fig. 9 we show  partial cross sections accounting 
for scalar and pseudoscalar exchanges (dashed curve)  and vector 
exchanges (dash-dotted curve) separately. The solid curve 
(labeled $\sigma^+$) gives the total cross section for the 
standard solution. Although the size of these partial cross sections 
are comparable, reversing the relative phases for both of the $\rho$ 
and $\omega$ to have the opposite signs hardly influences the 
predictions for the total cross section (curve labeled $\sigma^-$). 
Indeed, interference between vector mesons and scalar or pseudoscalar 
exchanges involves background terms only and
consequently are very small. Also, as we have   
indicated above the $\omega$ exchange is very weak in this case 
because N$^*$ pole terms sum to zero so that the main source of
ambiguities in the model predictions is due to 
interference amongst the scalar and pseudoscalar contributions. 

We obtain similar results for the $pn \rightarrow pn \eta$ reactions.
In this case contributions from both of 
the ${}^{33}P_0 \to {}^{31}\! S_0$ and ${}^{11}\!P_1 \to {}^{13}\!S_1$
transitions are possible and the cross section becomes
an incoherent sum,
\begin{equation}
\sigma_{np \to np \eta} = \frac {1}{2} \left[ 
\sigma_{11}\ +\ \sigma_{00} \right]~.
\label{eq:1}
\end{equation}
The calculated cross sections are shown in Fig. 10. 
Again the phase between the overall contribution from vector 
meson exchanges with respect to the overall contribution from
scalar and pseudoscalar does not affect much the calculated total 
cross section. In this case the process is dominated by a $\rho$
meson exchange. 
The contributions from $\omega$ exchanges through N$^*$ excitations
do not cancel out (see Eqn. 60), giving rise to a relatively 
large $\omega$  exchange contribution, so that interference with a
dominant $\rho$ exchange amplitude determines, to a large extent,  
the scale of the calculated cross section. As for the 
$pp \rightarrow pp \eta$ reaction, in the standard solution which 
explains data the $\rho$ and $\omega$ add destructively.
The isovector $\pi$ and $\delta$ contributions add destructively 
to those of the isoscalar $\eta$ and $\sigma$.

\section{FSI and comparison with data}
At energies close to threshold FSI corrections influence the 
energy dependence and scale of the calculated cross section. To 
allow comparison with data to be made, we introduce FSI in an 
approximate way following the procedure of Refs. 
\cite{moalem95,moalem96,hep95}. The S-wave transition 
amplitude for a three body reaction, say $NN \rightarrow NN \eta$, 
is assumed to factorize in the form
\begin{equation}
T_{if} \approx M_{if}^{in} T_{ff}^{el}~,
\end{equation}
where $M_{if}^{in}$ stands for the primary production amplitude 
and $T_{ff}^{el}$ represents the FSI correction factor. We 
identify this factor with the on mass-shell elastic scattering 
amplitude for the $NN \eta \rightarrow NN \eta$ (three particles 
in to three particles out transition).  The validity of this 
approximation was discussed at length previously\cite{hep95}. 
Here we recall only that, $T_{ff}^{el}$ is a solution of the 
Faddeev's equations and has the structure of the Faddeev's 
decomposition of the t-matrix for $3 \to 3$ transition. It is 
estimated using on mass shell two-body elastic scattering 
amplitudes, $t_{\eta N}$ and $t_{NN}$, in a three-body space 
where the third particle being a spectator. We 
calculate these from S-wave phase shifts. The S-wave $pp$ phase
shift is calculated using the modified Cini-Fubini-Stanghellini 
( CFS ) formula\cite{noyes72} with a proton-proton scattering 
length $a_{pp} = -7.82 fm$ and an effective range $r_{pp} = 2.7 fm$.
Similarly, the $np$ S-wave singlet and triplet phase shifts 
are calculated using the so called mixed effective range 
expansion with CFS shape corrections. The scattering lengths 
and effective ranges are taken to be $a_t = 5.413 $ fm and 
$r_t = 1.735 $ fm for triplet scattering and $a_s = -23.715 $ fm 
and $r_s = 2.73 $ fm for the singlet. The values of the CFS shape 
correction parameters are taken to be $p_1 = 0.1147 $ fm$^3$ 
and $p_2 = 3.861 $ fm$^2$. The $t_{\eta N}$ is taken from 
Ref. \cite{faldt95} to be
\begin{equation}
[t(\eta N \rightarrow \eta N)]^{-1} = 1/a + r k^2 / 2 - i k_{\eta}~,
\label{SL}
\end{equation}
with $\eta$p scattering length  and effective range
\begin{eqnarray} 
     a = (0.476 + 0.276 i) fm, &  & r = (-3.16 -0.13 i) fm~.
\label{SLV} 
\end{eqnarray}
Albeit, FSI corrections modify the energy dependence of the 
cross section at low energy region ($Q_{}$ below 15 MeV) 
and should not influence much predictions at higher energies. 
 
In Figs. 11-12 we draw our predictions for the energy 
integrated cross sections for the $pp \rightarrow pp \eta$ 
and $pn \rightarrow pn \eta$ reactions with FSI corrections
included, along with data taken from Refs. [2-6]. 
The solid curve gives the standard solution, i.e., with the 
relative phases of the various exchange amplitudes taken to be +1. 
The model explains fairly well the cross section data for 
both of these processes. Calculations which include resonance terms
only and without background terms are 
drawn as dashed curves. Clearly, the $\eta$ production is 
dominated by the N$^*$(1535 MeV) isobar excitations (diagram 2a). 
The effects of the background terms (diagrams 2b-2d) are 
bound to $10\%$ or 
less in the near threshold energy region considered here.
FSI corrections for the $np \rightarrow np \eta$ reaction are 
performed separately for $\sigma_{11}$ and $\sigma_{00}$ using the 
appropriate singlet and triplet $np$ S-wave phase shifts. 
The model reproduces the cross section data 
for all of the $pp \rightarrow pp \eta$, $pn \rightarrow pn \eta$
and $pn \rightarrow d \eta$ reactions. The results for the latter
process will be reported elsewhere\cite{moalem98}.

As indicated in the introduction, there exits no direct evidence 
for the coupling of 
the $\sigma$, $\delta$ and $\omega$ mesons to the N$^*$ (1535 MeV) 
isobar. It is interesting to see whether contributions from these
mesons are required to explain data. To this aim, calculations were 
repeated with : (i) $g_{\omega NN^*}$ = 0, (ii) $g_{\omega NN^*} = 
g_{\sigma NN^*} = 0$, (iii) $g_{\omega NN^*} = g_{\sigma NN^*} = 
g_{\delta NN^*} = 0$.
The predictions with $g_{\omega NN^*}$ = 0 are shown in Figs. 11-12
as dot-dashed curves. They explain equally well the data for the 
$pp \rightarrow pp \eta$ reaction, and for the $pn \rightarrow pn \eta$
reaction the agreement is even slightly better. The effects
of the  $\sigma$ and $\delta$ are negligibly small and their
contributions seem not to be needed either (see Figs. 12-13).

\begin{center}
\section{Summary and Discussion}
\end{center}
In this paper we have presented a covariant OBE model for $\eta$ 
production in NN collisions. The model includes nucleon and N$^*$
(1535 MeV) isobar excitations in the intermediate states and
contributions from all of the $\pi, \eta, \sigma, \rho, \omega$ and
$\delta$ meson exchanges. Our starting point was a Lagrangian 
which describes  NN scattering, in a consistent manner, in 
terms of these same meson exchanges.

As in other similar calculations, the model predictions are subject 
to uncertainties due to the unknown relative phases of the various
exchange amplitudes, but these are now limited to only a factor 
of 2-3  in the calculated cross sections.
The present formalism allows a thorough discussion of interference
between the various contributions. We have shown
that interference effects between vector and scalar or 
pseudoscalar meson exchanges are negligibly small.  Bearing
in mind that the most prominent contributions are those due to 
the $\pi$ and $\rho$ exchanges this feature of the model renders 
predictions to be considerably more reliable than our previous 
model calculations\cite{moalem96}. Assuming that the relative
phases of all of the various exchange amplitudes to be +1, the model
reproduces the cross section data, both the scale and energy
dependence, for the $pp \rightarrow pp \eta$ and
$pn \rightarrow pn \eta$ reactions. As indicated above the model 
also reproduces the cross section data for the 
$np \rightarrow d \eta$ reaction\cite{moalem98}

The $\omega$, $\sigma$ and $\delta$ 
contributions are small and with the present accuracies
of the measurements seem unnecessary to explain the data.  
This is not unwarranted property of the model since there 
is no direct evidence for 
the coupling of these mesons to the N$^*$ (1535 MeV) isobar.
Finally, with the $\rho$ coupling lowered  to $g_{\rho NN^*}$ = 0.615,
as low as suggested by Vetter et al.\cite{vetter91}, no solution is found
which explains data simultaneously for all of the $NN \rightarrow X\eta$
reactions.  It would be interesting to determine this coupling anew.\\

\vspace{2.0 cm}
{\bf Acknowledgments} This work was supported in part 
by the Israel Ministry Of Absorption.
We are indebted to  Z. Melamed
for assistance in computation. 

\newpage

\section{Appendix A} 

In this Appendix we describe the derivation of the expressions used 
to calculate $g_{\sigma NN^*}$ and $g_{\delta NN^*}$ (see caption of
Table 1).
Consider the effective triangle diagram of Fig. 14 and take 
the $\sigma \pi \pi$ vertex to have the general form,
\begin{equation}
V_{\sigma\pi\pi}=g_{\sigma\pi\pi}v_{\sigma\pi\pi}(k,q-k,q)~.
\label{vsigma}
\end{equation}
For a Lagrangian density,
$L_{int}=g_{\sigma\pi\pi}M\sigma\vec{\pi}\cdot\vec{\pi}$,
the function $v_{\sigma\pi\pi}(k,q-k,q)$  becomes,
\begin{equation}
v_{\sigma\pi\pi}(k,q-k,q) = M  f_1( k^2, q^2 )~,
\end{equation}
where $f_1(k^2,q^2)$ is a scalar function. Taking the  Lagrangian density to
be, $L_{int}= (g_{\sigma\pi\pi} / M )
(\sigma\partial\vec{\pi})\cdot(\partial\vec{\pi})$, 
yields,
\begin{equation}
v_{\sigma\pi\pi}(k,q-k,q)=\frac{1}{M}(q-k)q f_2(k^2,q^2)~.
\end{equation}
As will be demonstrated below the actual form of $v_{\sigma\pi\pi}$
does not influence the final results for the calculated coupling 
constants. Suppose now that one of the nucleon is on mass 
shell,$i.e.$, $p^2 = M^2$, then using the loop diagram Fig. 14
the $\sigma NN$ coupling is\cite{tiator94},
\begin{eqnarray}
   &  & g_{\sigma NN} v_{\sigma NN}(k,p-k,p) = -i 
\frac{ g_{\sigma\pi\pi}g^2_{\pi NN}}{(2\pi)^4}
\int d^4q { q\! \! \! / } v_{\sigma\pi\pi}(k,q-k,q)\nonumber \\
   &  & \left\{ (q^2-m^2_{\pi})[(q-k)^2-m^2_{\pi}]
 [(p-q)^2-M^2]\right\}^{-1}~.
\label{gsnn}
\end{eqnarray}
By using the standard Feynman parameterization and shift of 
the integration variable one obtains,
\begin{equation}
g_{\sigma NN}v_{\sigma NN}(k,p-k,p) = g_{\sigma\pi\pi}g^2_{\pi NN}
\left[p\! \! \! / X(k,p-k,p) + k\! \! \! / Y(k,p-k,p)\right]~, 
\label{gsnn1}
\end{equation}
where,
\begin{eqnarray}
 &  & X(k,p-k,p)=2\int dx_1dx_2dx_3 x_3 \delta(1-x_1-x_2-x_3))I_q~,\\
 &  & Y(k,p-k,p)=2\int dx_1dx_2dx_3 x_2 \delta(1-x_1-x_2-x_3))I_q~, 
\end{eqnarray}
and $I_q$ is the usual integral over the Euclidean four-vector 
$\tilde q$\cite{IZ},
\begin{eqnarray}
  &  & I_q=\frac{1}{(2\pi)^4}\int d^4{\tilde q}
           v_{\sigma\pi\pi}(k,q+px_3+k[x_2-1],q+px_3+kx_2)
\nonumber \\
  &  & ~~~~~~\left[({\tilde q}^2-(q+px_3+kx_2)^2+
              k^2x_2+m^2_{\pi}(x_1+x_3)\right]^{-3}~.
\end{eqnarray}
To evaluate the $g_{\sigma NN^*}$ coupling constant we have to replace one
of the two $\pi NN$ vertices with a $\pi NN^*$ vertex. This leads to,
\begin{equation}
g_{\sigma NN^*} v_{\sigma NN^*}(k,p-k,p) = 
g_{\sigma\pi\pi}g_{\pi NN}g_{\pi NN^*}
\left[p\! \! \! / X(k,p-k,p) + k\! \! \! / Y(k,p-k,p)\right]~. 
\label{gsnns1}
\end{equation}
After some algebra,
\begin{eqnarray}
 &  &  g_{\sigma NN}v_{\sigma NN}(k,p-k,p) = 
g_{\sigma\pi\pi}g^2_{\pi NN}J ~,\\
 &  & g_{\sigma NN*} v_{\sigma NN^*}(k,p-k,p) = 
g_{\sigma\pi\pi}g_{\pi NN}g_{\pi NN^*}J~,
\end{eqnarray}
with, 
\begin{equation}
J=\left[M^2X^2+k^2Y^2+2pkXY\right]^{1/2}~.
\end{equation}
From Eqns. 89 and 90, 
one sees that the functions $v_{\sigma NN}$ and $v_{\sigma NN^*}$
are identical up to a scale factor, which we may absorb into the constant
$g_{\sigma NN^*}$. Thus, Eqns. 89 and 90 are equivalent to,
\begin{equation}
\frac {g_{\sigma NN^*}}{g_{\sigma NN}} = 
                           \frac {g_{\pi NN^*} }{ g_{\pi NN}}
\label{scoupl} 
\end{equation} 
Note that taking the $\sigma$ coupling to be  
$i g_{\sigma NN^*} [ \bar{N^*} \gamma^5 N + \bar{N} \gamma^5 N^*]$, 
as in Eqn. \ref{lagran}, lead to a real $g_{\sigma NN^*}$. Otherwise 
with this coupling taken as
$g_{\sigma NN^*} [ \bar{N^*} \gamma^5 N - \bar{N} \gamma^5 N^*]$,
one obtains a pure imaginary $g_{\sigma NN^*}$. 

Similar calculations can be made for the $g_{\delta NN^*}$ leading to
a similar expression, $viz$,
\begin{equation}
\frac {g_{\delta NN^*}}{g_{\delta NN}} = \frac {1}{2} \left[
                           \frac {g_{\pi NN^*} }{ g_{\pi NN}} + 
                           \frac {g_{\eta NN^*} }{ g_{\eta NN}}
\right]~.
\label{scoupl} 
\end{equation}

\section{Appendix B}
In what follows we write explicit expressions for the vectors and 
tensors of Eqns. 14-16. Let $i, j$ label the i-th and j-th nucleon
(see Fig. 1). We define,
\begin{eqnarray}
 & & \phi_{ij}^{\pm} = 1 \pm \Pi_i \cdot \Pi_j~,\\
 & & \Sigma_{ij} = \Pi_i + \Pi_j~,\\
 & & \Delta_{ij} = \Pi_i - \Pi_j~,\\
 & & \chi_{ij} = \Pi_i \times \Pi_j~.
\end{eqnarray}
Then for a scalar  meson exchange,
\begin{eqnarray}
 & & S_0 =  i F_S(q^2) G_S(q^2) T_S^{(1)}~, \\
 & & S_{13}^l = F_S(q^2) G_S(q^2) T_S^{(1)} \chi_{13}^l ~,\\ 
 & & \tilde{S}_{24}^l = F_S(q^2) G_S(q^2) \phi_{13}^- T_S^{(2)}~,\\
 & & {S}^{kl} = i F_S(q^2) G_S(q^2) \chi_{13}^k  T_S^{(2)l}~,
\end{eqnarray}
with,
\begin{eqnarray}
 & & T_S^{(1)} = \frac {1}{2} ({\bf k} + {\bf q})\cdot \chi_{13} B_S~,\\
 & & T_S^{(2)} = A_S \Delta_{24}^l + \frac{1}{2} B_S [
({\bf k} + {\bf q})\cdot \Pi_4 \Pi_2^l + 
({\bf k} + {\bf q})\cdot \Pi_2 \Pi_4^l + \nonumber \\
 & & ~~~~~~~~~\phi_{24}^- ({\bf k} + {\bf q})^l - 
(k_0 + q_0) \Sigma_{24}^l ] ~.
\end{eqnarray}
For a pseudoscalar exchange,
\begin{eqnarray}
 & & P_{13}^l = G_P(q^2) F_P(q^2) T_P^{(1)}
\Delta_{13}^l~,\nonumber \\ 
 & & {P}^{kl} = i F_P(q^2) G_P(q^2) \Delta_{13}^k  T_S^{(2)l}~, 
\end{eqnarray}
with,
\begin{eqnarray}
 & & T_P^{(1)} = A_P \phi_{24}^- + \frac{1}{2} B_S [
\phi_{24}^+( k_0 + q_0) + 
({\bf k} + {\bf q})\cdot \Sigma_{24} ]~,\nonumber \\
 & & T_P^{(2)l} = A_P \chi_{24}^l - \frac{1}{2} B_S [
(k_0 + q_0) \chi_{24}^l + 
({\bf k} + {\bf q})\times \Delta_{24}^l ]~.
\end{eqnarray} 
For a vector meson exchange,
\begin{eqnarray}
 & & V_0 =  i G_V(q^2) C_{13}^{\mu} X_{24,\mu}~, \nonumber \\
 & & V_{13}^l = - G_V(q^2) D_{13}^{\mu l} X_{24,\mu}~, \nonumber \\
 & & \tilde{V}_{24}^l = G_V(q^2) C_{13}^{\mu l} Y_{24,\mu}^l~,\nonumber \\
 & & V^{ln} = i G_V(q^2) D_{13}^{\mu l} Y_{24,\mu}^n~,  
\end{eqnarray}
where,
\begin{equation}
C_{13}^0 = F_V^{(1)} \phi_{24}^+  +  F_V^{(2)} {\bf q} \cdot
\Delta_{13} ~,
\end{equation}
\begin{equation}
C_{13}^m = F_V^{(1)} \Sigma_{24}^m +  F_V^{(2)} (q^0 \Delta_{13}^m
 - \Pi_1^m {\bf q}\cdot\Pi_3 - \Pi_3^m {\bf q}\cdot\Pi_1)~,
\end{equation}
\begin{equation}
D_{13}^{0k} = -\chi_{13}^k F_V^{(1)} + 
{\bf q} \times \Sigma_{13}^k F_V^{(2)}~,
\end{equation}
\begin{equation}
D_{13}^{mk} = \epsilon^{mrs} [ (\Delta_{13}^r F_V^{(1)} +
q^0 \Sigma_{13}^r + 
\phi_{13}^+ q^r F_V^{(2)})\delta^{sk} 
- q^r(\Pi_3^s \Pi_1^k + \Pi_1^s \Pi_3^k) F_V^{(2)} ]~,
\end{equation}
\begin{eqnarray}
 & &  X_{24}^0 = {\bf k} \cdot \chi_{24}~ a_{46}^0~,\\
 & &  X_{24}^m = {\bf k} \cdot \chi_{24}~ a_{46}^m 
- \epsilon^{mrs} \Pi_2^r \Pi_4^s A_1 - \epsilon^{mrs} k^r 
\Delta_{24}^s A_2~,
\end{eqnarray}
\begin{eqnarray}
 & & Y_{24}^{0,l} = (a_{35}^0 + k^0 A_2)\Delta_{24}^l - 
     a_{46}^0 [k^0 \Sigma_{24}^l - \phi_{24}^- k^l + 
\Pi_2^l \Pi_4\cdot {\bf k} - \Pi_4^l \Pi_2\cdot {\bf k} ]~~~~~~~ 
\nonumber \\
 & & ~~~~~~~~~~~~~~~~~~~~~ - A_1 \Sigma_{24}^l + A_2 [k^l\phi_{24}^- 
+ \Pi_2^l \Pi_4\cdot {\bf k} + \Pi_4^l \Pi_2\cdot {\bf k} ]~~,\nonumber \\
 & & Y_{24}^{m,l} = (a_{35}^m + k^m A_2)\Delta_{24}^l - 
     a_{46}^m (k^0 \Sigma_{24}^l - \phi_{24}^-k^l
     - \Pi_2^l \Pi_4\cdot {\bf k} - \Pi_4^l \Pi_2\cdot {\bf k})~~~~~~ 
\nonumber \\
 & & ~~~~~~~~~~~~~~~~~~~~~ - A_1 (\phi_{24}^- \delta^{ml} + \Pi_4^m \Pi_2^l + 
\Pi_2^m \Pi_4^l ) 
     - A_2 [ {\bf k} \cdot \Delta_{24} \delta^{ml} 
     -\Delta_{24}^m k^l ~~\nonumber \\
 & & ~~~~~~~~~~~~~~~~~~~~~ - k^0 ( \phi_{24}^- \delta^{ml} + \Pi_4^m \Pi_2^l + 
       \Pi_2^m \Pi_4^l )]~. 
\end{eqnarray}

\newpage
\begin{table}
\centering
\caption{The model parameters. The constants
$g_{BNN}$, cut off parameters ($\Lambda_B$) and meson masses ($m_B$)
are taken from Machleidt et al.\cite{machleidt89}; their 
relativistic OBEP solution 
with pseudoscalar coupling for $\pi$ and $\eta$ and with 
$f_{\rho NN}/g_{\rho NN} = 6.1, f_{\omega NN}/g_{\omega NN} = 0$~.}
\vskip 0.1 in
\begin{tabular}{|l|l|l|l|l|}
\hline
$$ & $m_B (MeV)$ & $g_{BNN}^2/4\pi$ & $\Lambda_B (MeV)$ & $g_{BNN^*}$\cr
\hline
 $\pi    $ & $ 138         $ & $ 14.6    $ & $ 1300   $ & $ 0.8   $\cr
 $\sigma $ & $ 550         $ & $  8.03   $ & $ 1800   $ & $ 0.5^{b)} $\cr
 $\eta   $ & $ 547.3^{a)} $ & $  3.0    $ & $ 1500   $ & $ 2.2   $\cr
 $\rho   $ & $ 769         $ & $  0.95   $ & $ 1300   $ & $ 1.66  $\cr
 $\omega $ & $ 783         $ & $ 20.0    $ & $ 1500   $ & $ 0.94^{b)}$\cr
 $\delta $ & $ 983         $ & $  5.07   $ & $ 1500   $ & $ 1.48^{c)}   $\cr
\hline
\end{tabular}
\ \\
\begin{flushleft}
~~~~~~~~~~~~~~~~~~~~~~~~~\small {${}^{a)}$ From Ref.\cite{plouin90}}~. \\
~~~~~~~~~~~~~~~~~~~~~~~~~~~\small {${}^{b)}$ From
   $g_{\sigma NN^*} /g_{\sigma NN} = g_{\pi NN^*} / g_{\pi NN}$} \\
~~~~~~~~~~~~~~~~~~~~~~~~~~~\small {${}^{c)}$ From
   $g_{\delta NN^*} /g_{\delta NN} = \frac {1}{2}\left[
  g_{\pi NN^*} / g_{\pi NN} + g_{\eta NN^*} / g_{\eta NN}\right]$} \\
\end{flushleft}
\label{mesonpar}
\end{table}

\setlength{\oddsidemargin}{0in}
\setlength{\evensidemargin}{1.6in}
\setlength{\textwidth}{6.0in}
\setlength{\topmargin}{-.6in}
\setlength{\textheight}{9.0in}
\pagestyle{plain}
\setlength{\baselineskip} {4.0ex}

\newpage
\vspace{0.4cm}
\begin{figure}
\vspace{4.0in}
\includegraphics{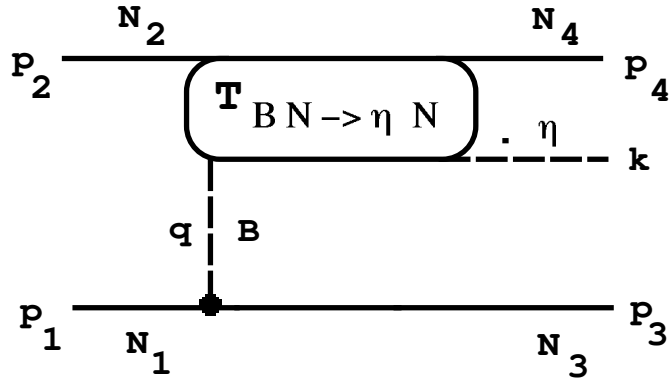}
\vskip 0.5 in
\caption{ The primary production mechanism for the 
$NN \to NN \eta$ reaction. A boson B created on nucleon 1 (momentum 
$p_1$), is converted into an $\eta$ meson (momentum k) on nucleon 2
(momentum $p_2$).  
}  
\end{figure}

\vspace{0.4cm}
\begin{figure}
\vspace{3.0in}
\includegraphics{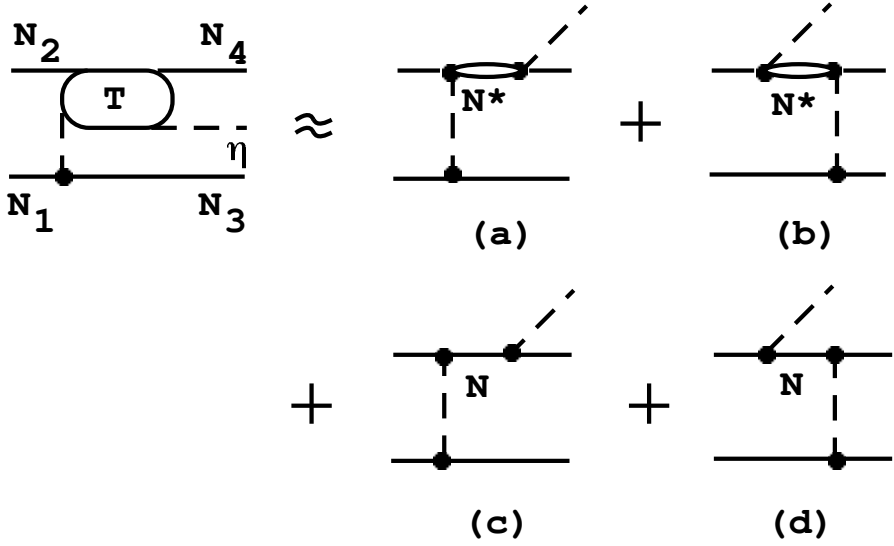}
\vskip 0.5 in
\caption{  Feynman diagrams for $\eta$ meson production in NN 
collisions. (a) s-channel N$^*$ pole contribution; a boson B (dashed line) 
formed on nucleon 1 (momentum $p_1$) is absorbed on nucleon 2 
(momentum $p_2$) which is excited to an isobar state 
with momentum $p=k+p_4$ which then decays into a nucleon 4 
(momentum $p_4$) and an $\eta$ meson (momentum k). (b) u-channel 
N$^*$ pole; pre-emission counter part
of diagram (a) where an $\eta$ meson is emitted before
collision. (c) s-channel N pole; as in (a) with a nucleon in the 
intermediate state. (d) u-channel N pole contribution; pre-emission
counter part of (c). There are four more graphs in which the exchanged 
boson is formed on nucleon 2. These are not displayed here but their
contributions are included in the calcultions.
}  
\end{figure}

\vspace{0.4cm}
\begin{figure}
\vspace{3.0in}
\includegraphics{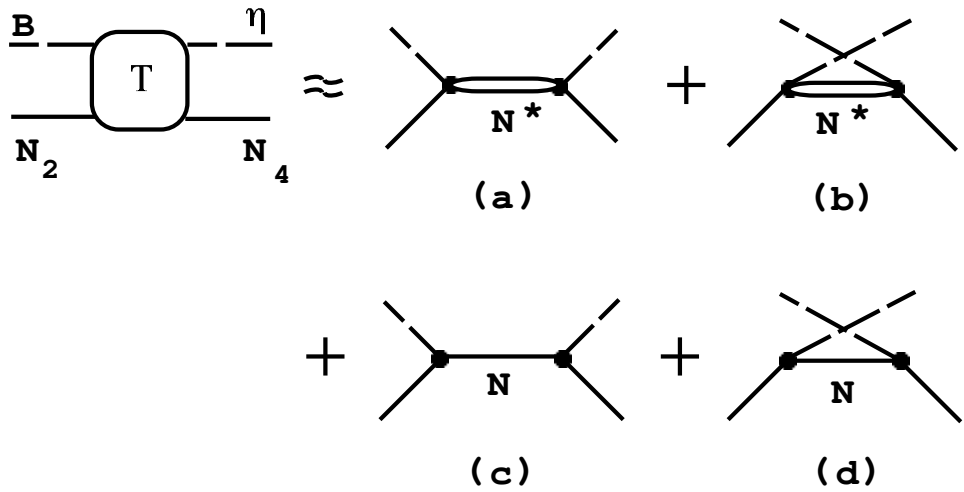}
\vskip 0.5 in
\caption{ Feynman diagrams for the conversion process 
$BN \rightarrow \eta N$ 
(a) s-channel N$^*$ pole contribution, (b) u-channel N$^*$ pole; pre-emission
counter part of diagram (a), (c) s-channel N pole; as in (a) with a nucleon 
in the intermediate state. (d) u-channel N pole contribution; pre-emission
counter part of (c). 
}  
\end{figure}

\begin{figure}[t]
\vspace{5.0in}
\includegraphics{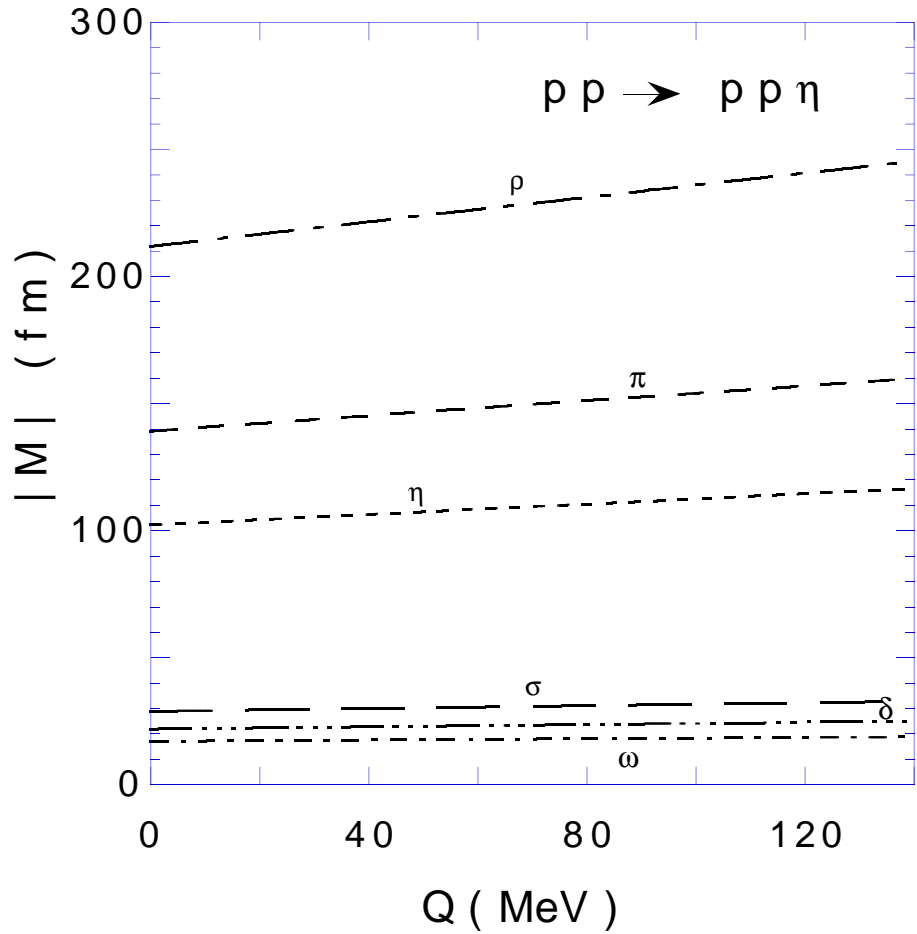}
\vskip 0.5 in
 \caption{  Partial exchange amplitude for the $pp \to pp \eta$
reaction. The $\omega$ contributions from s and u-channels 
with N$^*$ poles cancel out and $M_{\omega}$ is very weak (see text).
}
    \label{fig3}
\end{figure}

\begin{figure}[t]
\vspace{4.5in}
\includegraphics{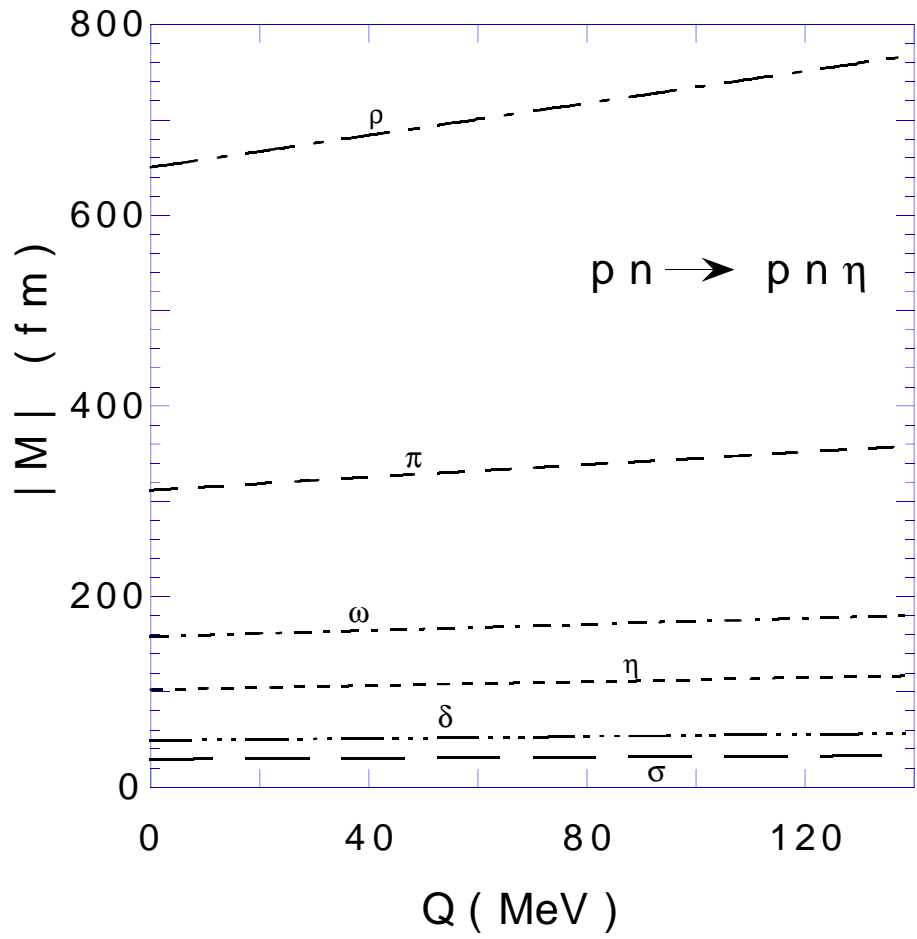}
\vskip 0.5 in
 \caption{  Partial exchange amplitude for the $pn \to pn \eta$
reaction. 
}
    \label{fig4}
\end{figure}

\begin{figure}[t]
\vspace{5.0in}
\includegraphics{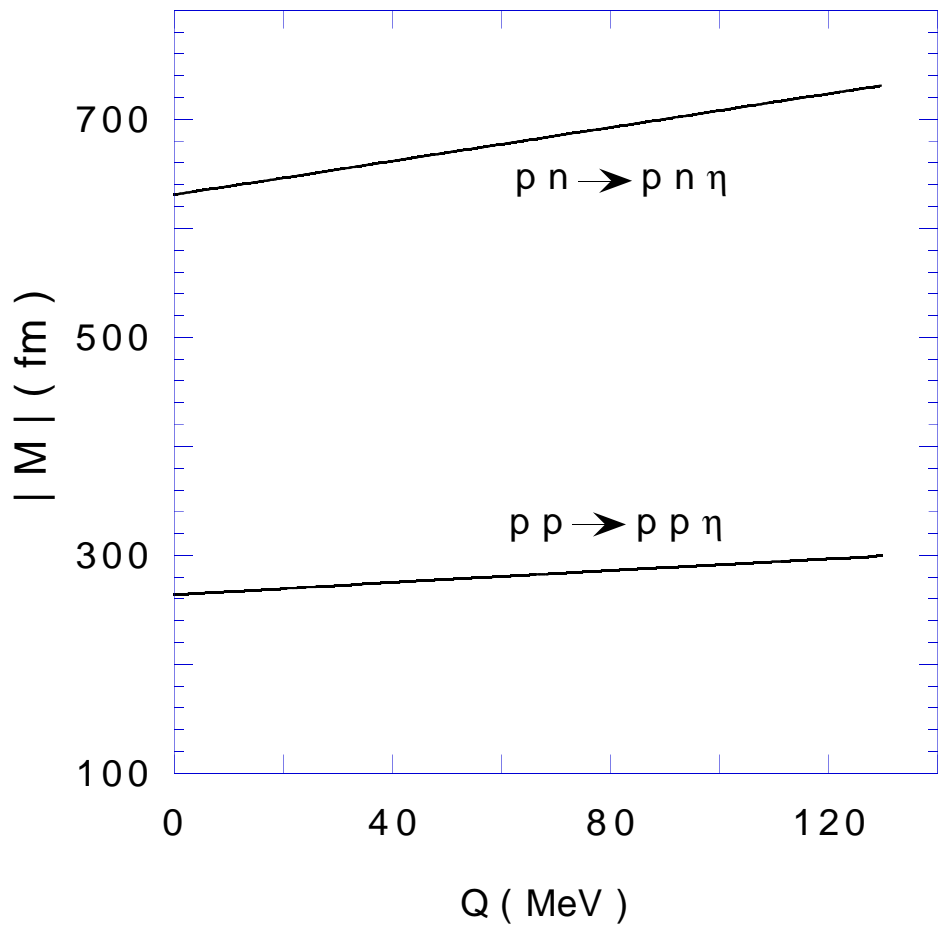}
\vskip 0.5 in
 \caption{ Primary production amplitude for the 
$pp \rightarrow pp \eta$ and $pn \rightarrow pn\eta$ reactions
with all of the relative phases taken to be +1 (our standard solution).
}
    \label{fig14}
\end{figure}

\begin{figure}[t]
\vspace{5.0in}
\includegraphics{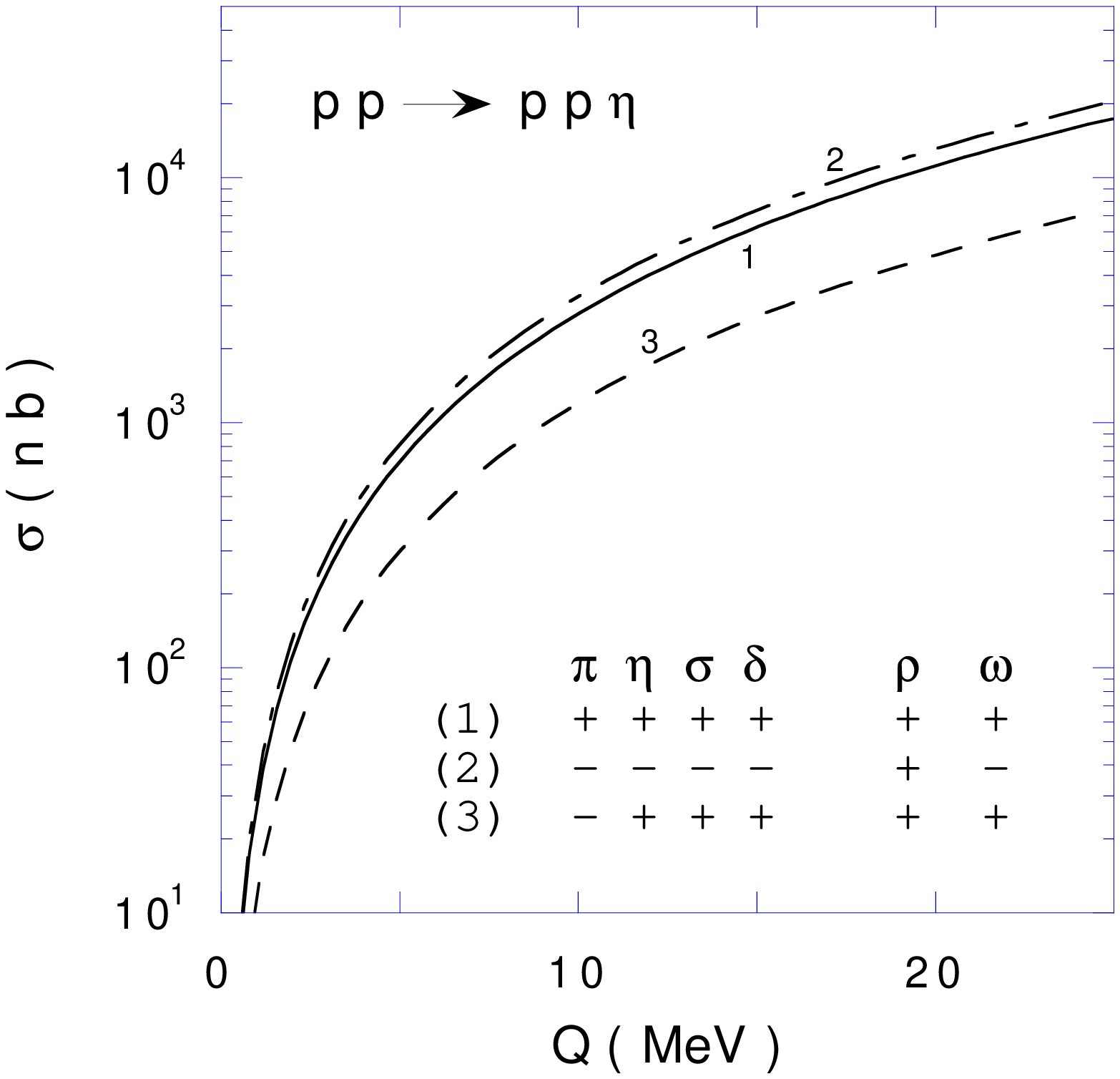}
\vskip 0.5 in
 \caption{ Predictions for the  $pp \to pp \eta$ cross sections near
threshold with different phase combinations. The solid line displays
the standard solution. The lowest (dashed curve) and the highest
(dash-dotted curve) cross sections are obtained with the $\pi, \eta,
\sigma, \delta, \rho$ and $\omega$ relative phases being $-+++++$ and
$----+-$, respectively. 
}
    \label{fig5}
\end{figure}

\begin{figure}[t]
\vspace{5.0in}
\includegraphics{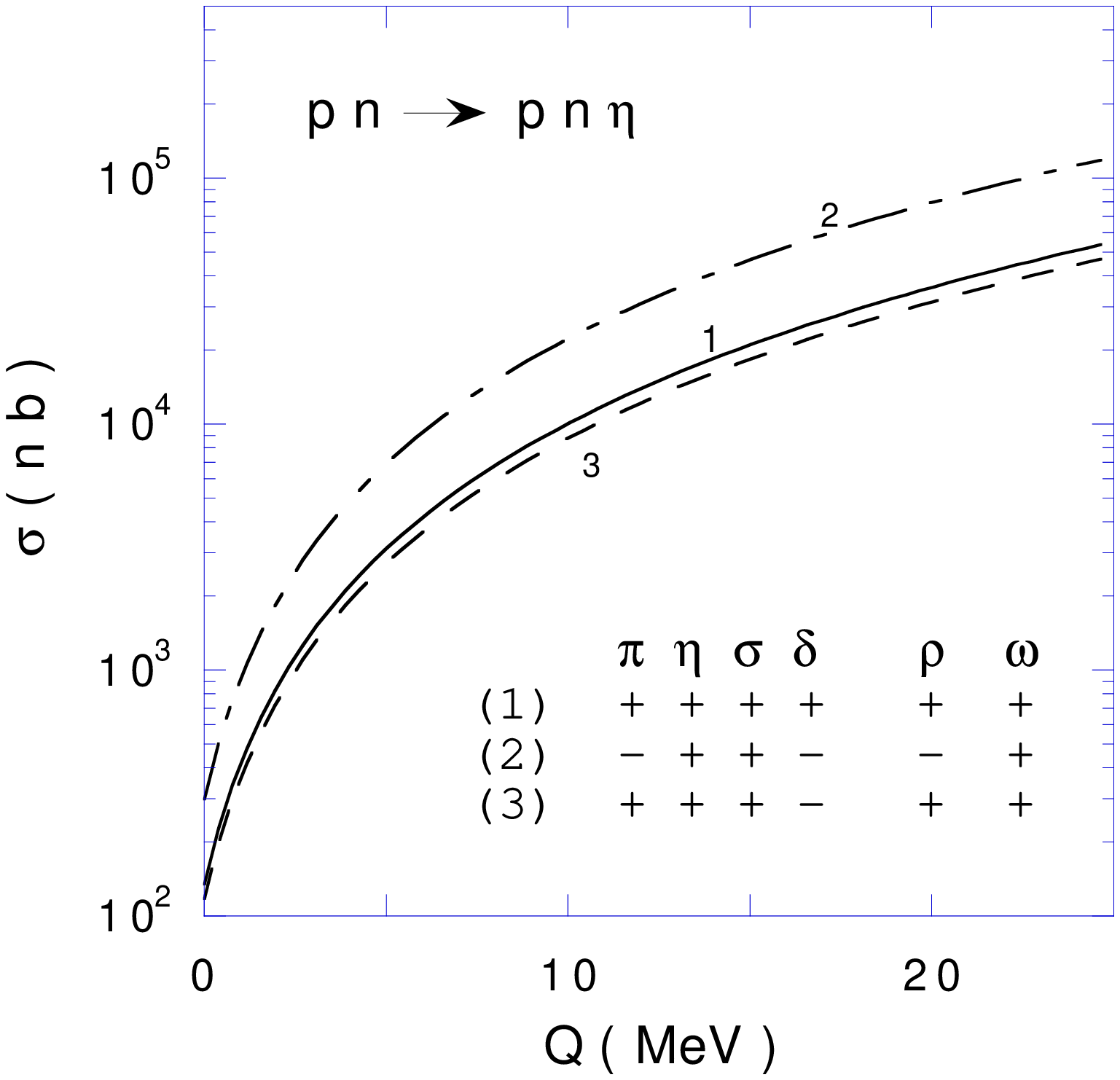}
\vskip 0.5 in
 \caption{ Predictions for the  $pn \to pn \eta$ cross sections near
threshold with different phase combinations. The solid line displays
the standard solution. The lowest (dashed curve) and the highest
(dash-dotted curve) cross sections are obtained with the $\pi, \eta,
\sigma, \delta, \rho$ and $\omega$ relative phases being $+++-++$ and
$-++--+$, respectively. 
}
    \label{fig6}
\end{figure}

\begin{figure}[t]
\vspace{5.0in}
\includegraphics{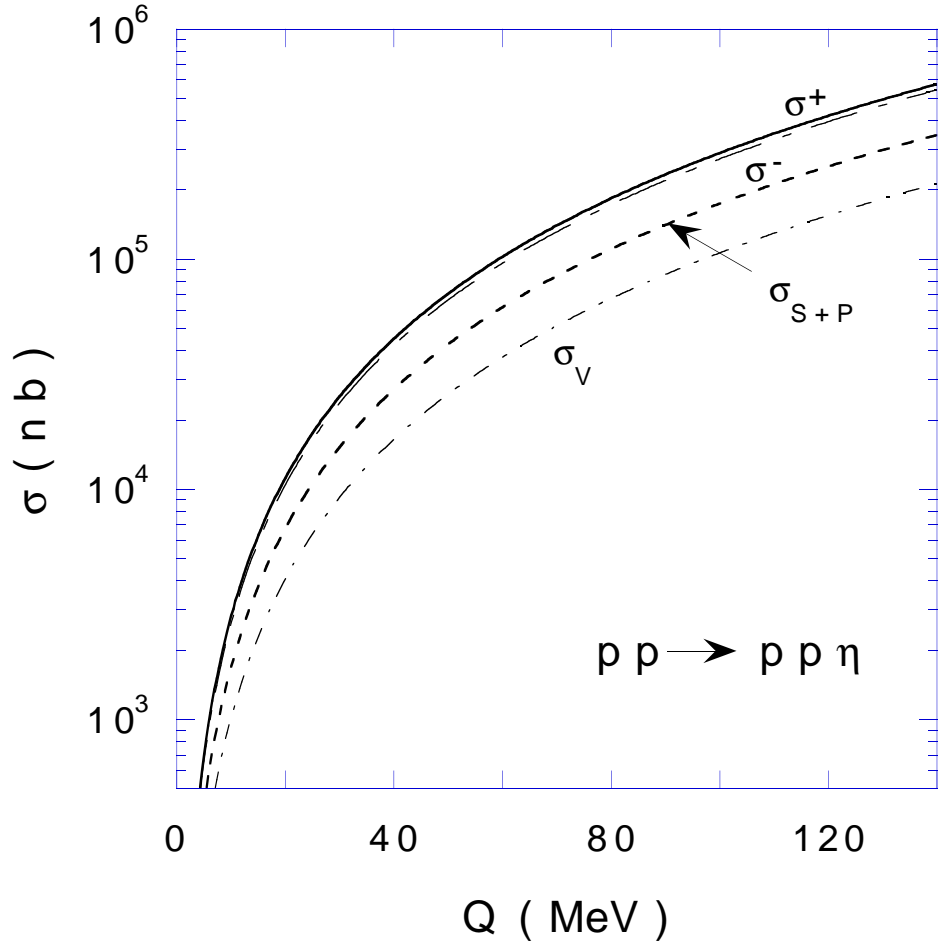}
\vskip 0.5 in
 \caption{ Interference between vector and  scalar-pseudoscalar meson
contributions. The curves labeled by $\sigma_V , \sigma_{P+S},$
represent partial cross sections with vector meson 
exchanges and  with scalar plus pseudoscalar exchanges only.
The other lines are total cross sections with all the relative phases
taken as +1 ($\sigma^+$) and with the same relative phases for the 
scalars and pseudoscalars except for $M_{\rho}$ and $M_{\omega}$ 
having the oposite sign ($\sigma^-$).
}
    \label{fig7}
\end{figure}

\begin{figure}[t]
\vspace{5.0in}
\includegraphics{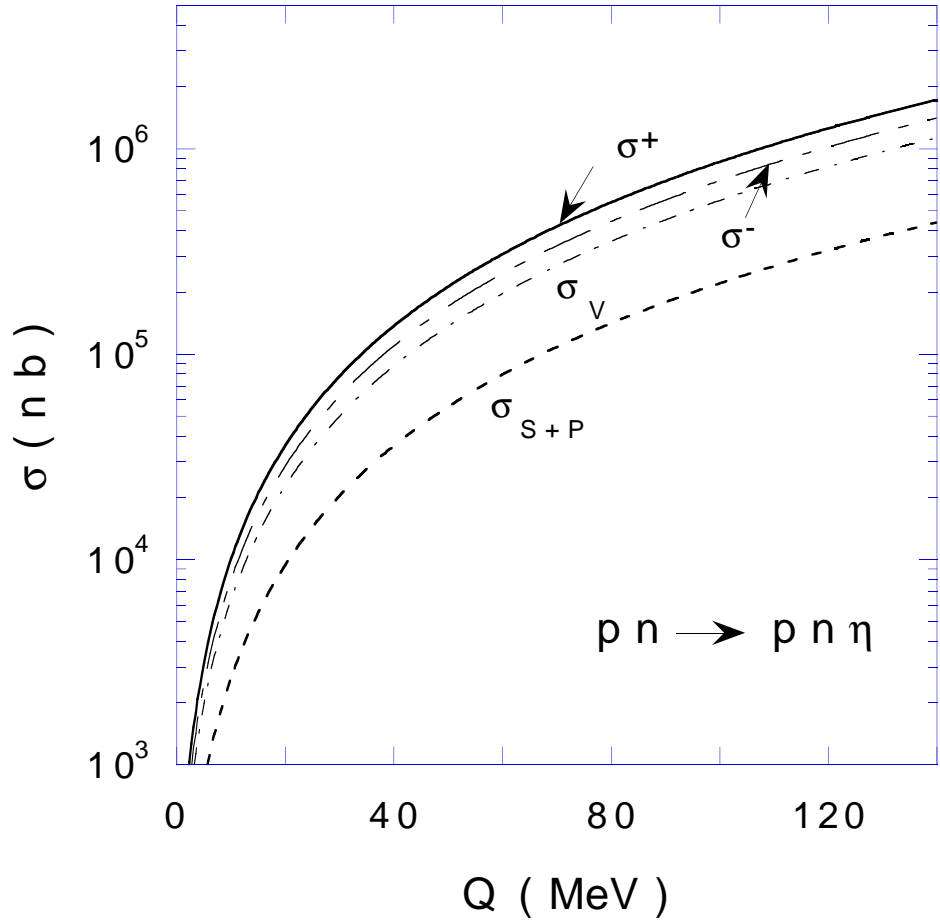}
\vskip 0.5 in
 \caption{ Interference between vector and  scalar-pseudoscalar meson
contributions for the $pn \rightarrow pn \eta$ reaction. See captions 
of Fig. 9.
}
    \label{fig8}
\end{figure}

\begin{figure}[t]
\vspace{5.5in}
\includegraphics{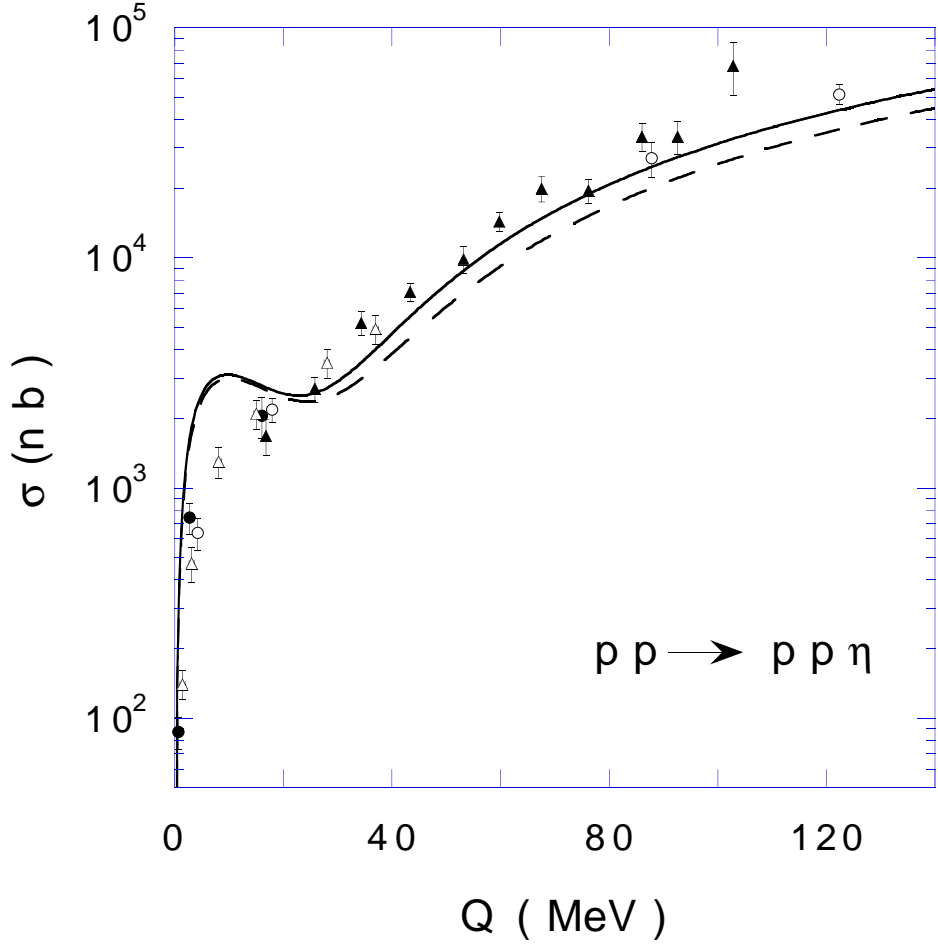}
\vskip 0.5 in
 \caption{ Energy integrated cross sections for the $pp \rightarrow pp \eta$
reaction $versus$ the energy available in the CM system. All curves include 
FSI corrections via the approximation Eqn. 77. Predictions
are given for the standard solution with both resonance and background 
terms included (solid line) and with the resonance term only
(dashed curve).  
The dot-dashed curve (which in this case override the solid line) 
gives the standard solutions as obtained with $g_{\omega} =0$.
The data shown are taken from Refs. 2-4,7.
}    

\end{figure}

\begin{figure}[t]
\vspace{5.5in}
\includegraphics{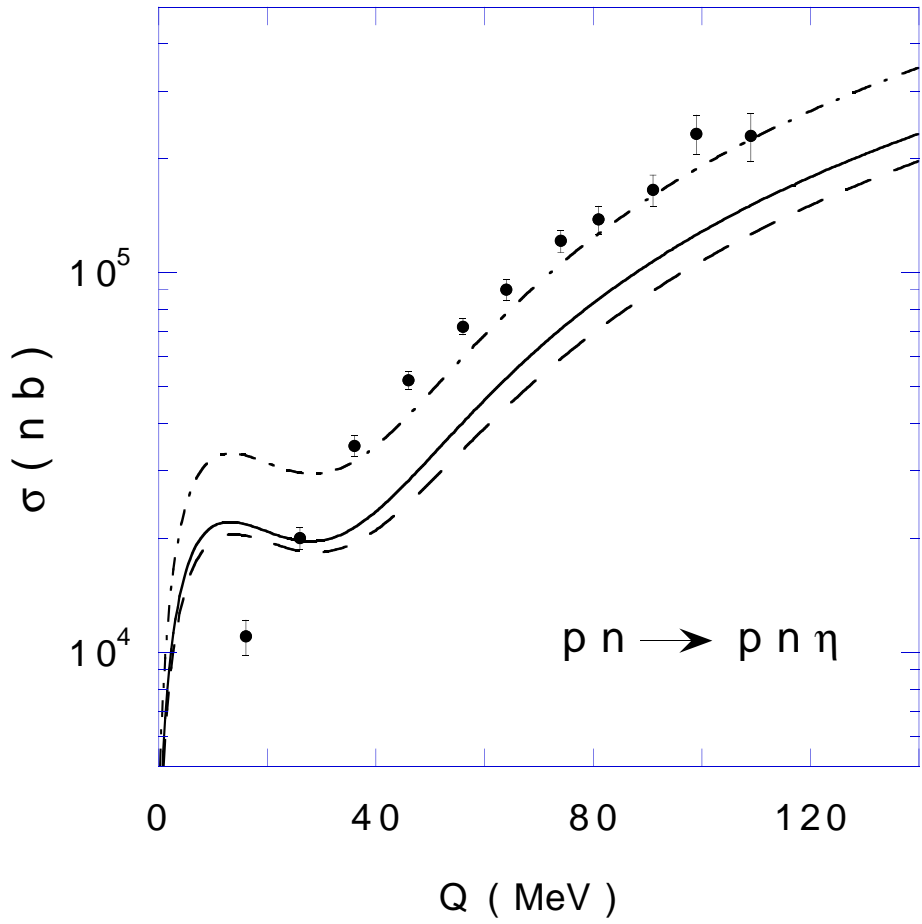}
\vskip 0.5 in
 \caption{ Energy integrated cross section for the $pn \rightarrow pn \eta$
reaction. Data point are taken from Ref. 7. See caption of Fig. 11.
}    
\end{figure}

\begin{figure}[t]
\vspace{5.0in}
\includegraphics{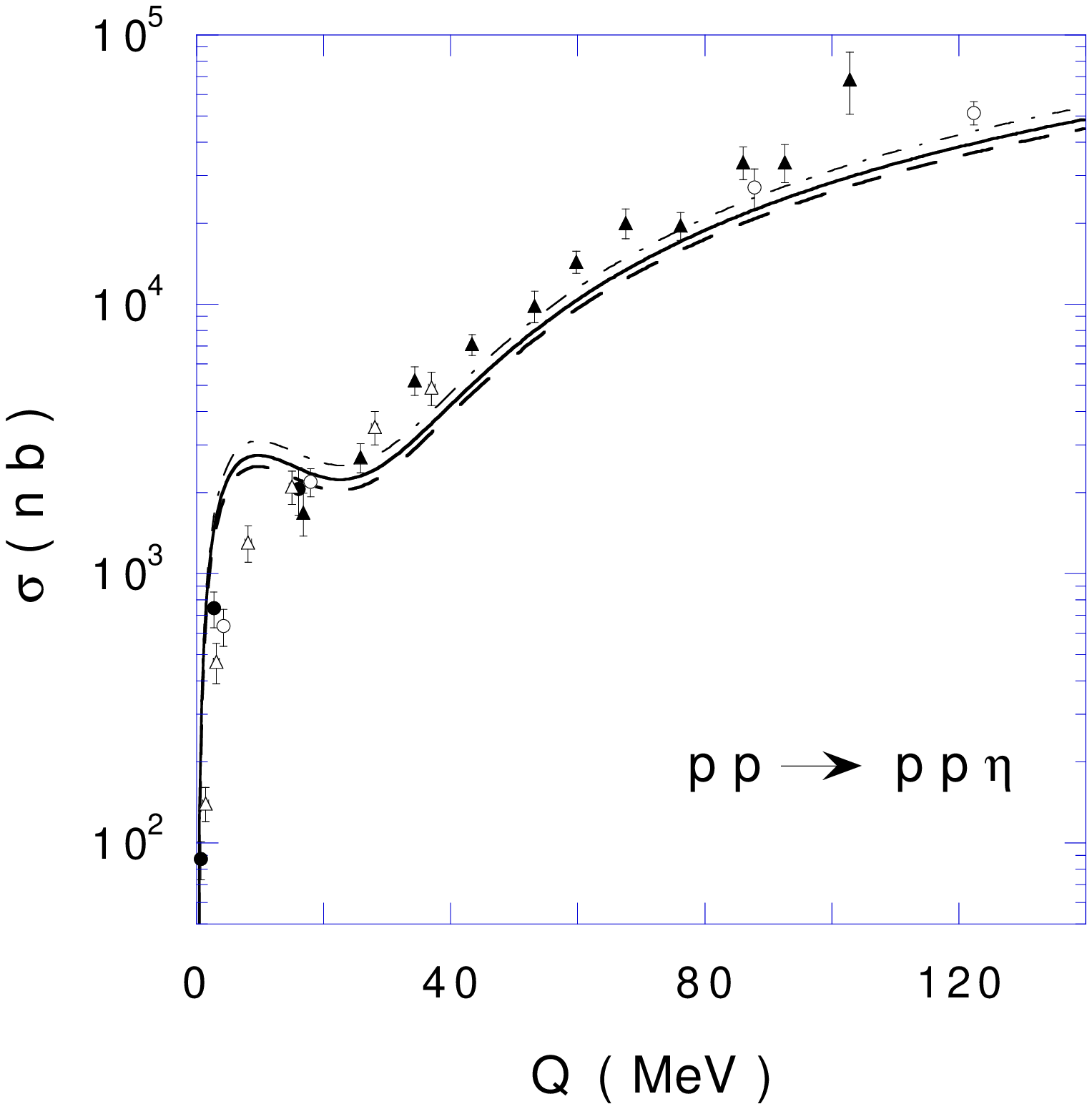}
\vskip 0.5 in
 \caption{ Effects of $\omega, \sigma, \delta$ exchanges. 
Predictions for the $pp \rightarrow pp \eta$ cross section are given 
for the standard solution with 
(i) $g_{\omega NN^*}=0$ (solid line), 
(ii)$ g_{\omega NN^*} = g_{\sigma NN^*} =0$ (dashed curve) and 
(iii) $g_{\omega NN^*} = g_{\sigma NN^*} = g_{\delta NN^*} = 0$ 
(dash-dotted curve).
}
    \label{fig11}
\end{figure}

\begin{figure}[t]
\vspace{5.0in}
\includegraphics{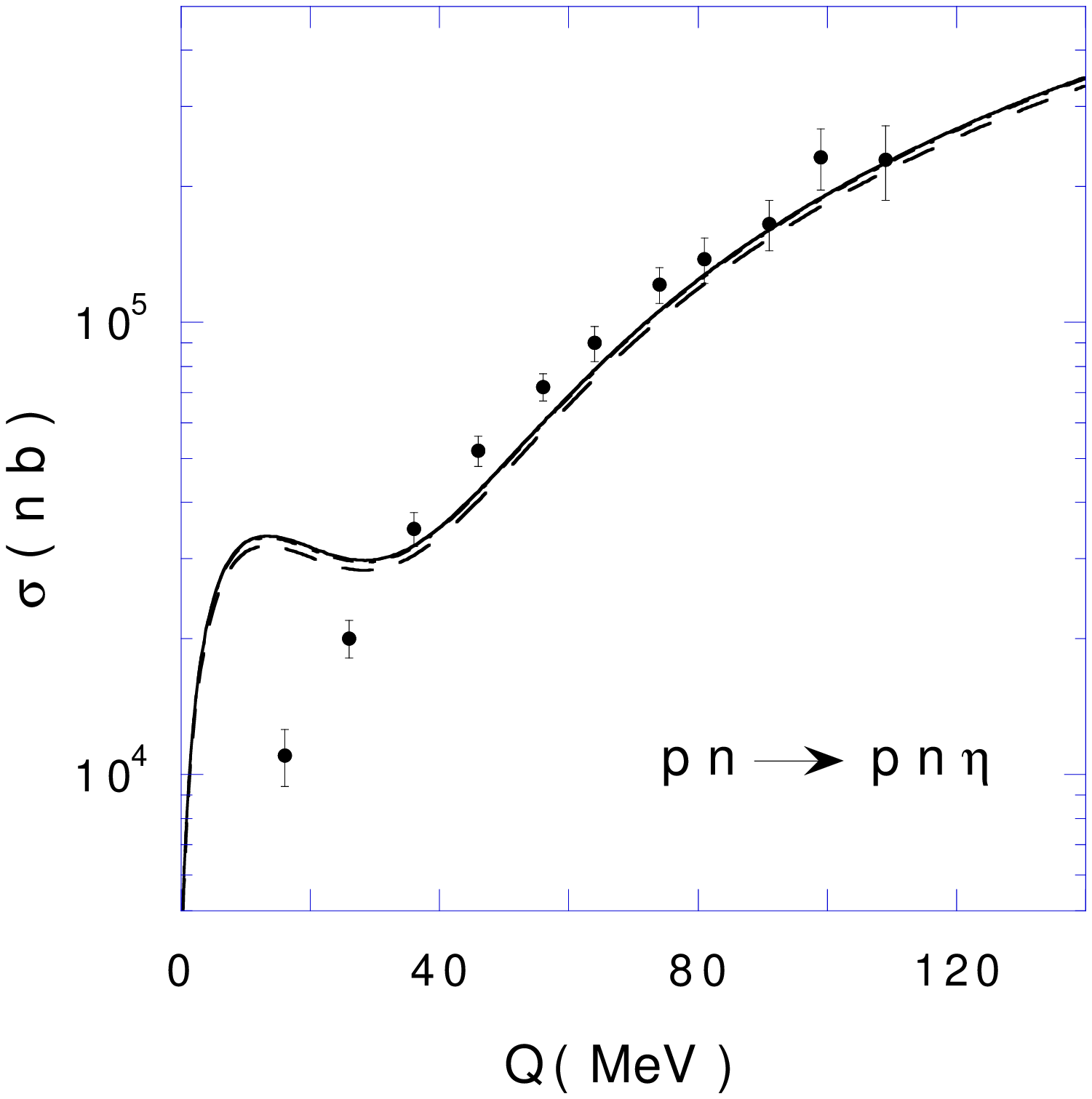}
\vskip 0.5 in
 \caption{ Effects of $\omega, \sigma, \delta$ exchanges. 
Predictions for the $pn \rightarrow pn \eta$ cross section are given 
for the standard solution with 
(i) $g_{\omega NN^*}=0$ (solid line), 
(ii)$ g_{\omega NN^*} = g_{\sigma NN^*} =0$ (dashed curve) and 
(iii) $g_{\omega NN^*} = g_{\sigma NN^*} = g_{\delta NN^*} = 0$ 
(dash-dotted curve).
}
    \label{fig12}
\end{figure}

\begin{figure}[t]
\vspace{5.0in}
\includegraphics{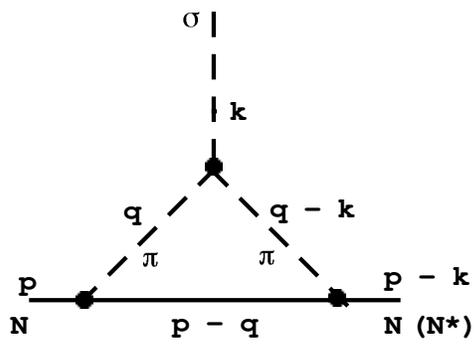}
\vskip 0.5 in
 \caption{ An effective triangle diagram used to derive an expression
for $g_{\sigma NN^*}$
}
    \label{fig13}
\end{figure}

\end{document}